\documentclass[aps,pre,superscriptaddress,floatfix,amsmath,amssymb,reprint]{revtex4-2}

\pdfoutput=1

\usepackage{color}
\usepackage{frcursive}
\usepackage{bbm}
\usepackage{mathtools}
\usepackage{graphicx}
\usepackage{enumitem}
\usepackage{bbm}
\usepackage{relsize}
\usepackage{xcolor}
\usepackage[caption=false]{subfig}
\usepackage{amsmath}	
\usepackage{amssymb}    
\usepackage[pdftex]{hyperref}

\newtheorem{theo}{Theorem}[section]

\newtheorem{prop}{Proposition}[section]

\newtheorem{definition}{Definition}[section]

\let \leq \leqslant
\let \geq \geqslant

\let \epsilon \varepsilon

{ \par \medskip \par
  \noindent \textit{\textbf{Demonstration\/}} : }{\null \hfill $\Box$ \par }
 
\newcommand{\R} {\ensuremath{\mathbb{R}}}
{

\newcommand{\C} {\ensuremath{\mathbb{C}}}

\begin{document}

\title{Numerical quasi-conformal transformations for electron dynamics on strained graphene surfaces}

\author{Fran\c{c}ois Fillion-Gourdeau}
\email{francois.fillion@emt.inrs.ca}
\affiliation{Institute for Quantum Computing, University of Waterloo, Waterloo, Ontario, Canada, N2L 3G1}
\affiliation{Infinite Potential Laboratories, Waterloo, Ontario, Canada, N2L 0A9}

\author{Emmanuel Lorin}
\email{elorin@math.carleton.ca}
\affiliation{School of Mathematics and Statistics, Carleton University, Ottawa, Canada}
\affiliation{Centre de Recherches Math\'ematiques, Universit\'e de Montr\'eal, Montr\'eal, Canada.}

\author{Steve MacLean}
\email{steve.maclean@emt.inrs.ca}

\affiliation{Institute for Quantum Computing, University of Waterloo, Waterloo, Ontario, Canada, N2L 3G1}
\affiliation{Infinite Potential Laboratories, Waterloo, Ontario, Canada, N2L 0A9}
\affiliation{Universit\'{e} du Qu\'{e}bec, INRS-\'{E}nergie, Mat\'{e}riaux et T\'{e}l\'{e}communications, Varennes, Qu\'{e}bec, Canada J3X 1S2}

\date{\today}

\date{\today}

\begin{abstract}
The dynamics of low energy electrons in general static strained graphene surface is modelled mathematically by the Dirac equation in curved space-time. In Cartesian coordinates, a parametrization of the surface can be straightforwardly obtained, but the resulting Dirac equation is intricate for general surface deformations. Two different strategies are introduced to simplify this problem: the diagonal metric approximation and the change of variables to isothermal coordinates. These coordinates are obtained from quasi-conformal transformations characterized by the Beltrami equation, whose solution gives the mapping between both coordinate systems. To implement this second strategy, a least square finite-element numerical scheme is introduced to solve the Beltrami equation. The Dirac equation is then solved via an accurate pseudo-spectral numerical method in the pseudo-Hermitian representation that is endowed with explicit unitary evolution and conservation of the norm. The two approaches are compared and applied to the scattering of electrons on Gaussian shaped graphene surface deformations. It is demonstrated that electron wave packets can be focused by these local strained regions.   
\end{abstract}

\maketitle

\section{Introduction}

Graphene is a material made of carbon atoms arranged on a 2D honeycomb lattice. This atomic configuration confers this material with unique electronic, mechanical and thermal properties \cite{doi:10.1142/9789814287005_0002}. In particular, its charge carriers can be described by a massless Dirac equation at low energy ($\lesssim 2$ eV), promoting graphene to the rank of Dirac materials \cite{doi:10.1080/00018732.2014.927109}. Thanks to these interesting and intriguing properties, graphene has been considered for many applications, ranging from quantum electrodynamics simulators \cite{KATSNELSON20073} to field effect transistors \cite{doi:10.1063/1.3077021}. 

A little more than a decade ago, it has been realized that electronic properties of charge carriers can be modified by the introduction of mechanical deformations (strain) in graphene samples \cite{Cortijo_2007,PhysRevB.76.165409}. This gave birth to the field of strain engineering or ``straintronics'' \cite{PhysRevLett.103.046801,PhysRevB.81.081407,GUINEA20121437,AMORIM20161,Feng2016,Naumis_2017}. This phenomenon has spurred many theoretical \cite{PhysRevLett.108.227205,PhysRevB.88.085430,PhysRevB.88.155405,RAMEZANIMASIR201376,PhysRevB.95.125432,PhysRevB.84.081401} and experimental investigations \cite{PhysRevB.81.035408,doi:10.1021/nn800459e,Levy544,Klimov1557}, in part because strain is responsible for very large pseudo-magnetic fields and because straintronics may pave the way towards technological and scientific applications. For instance, the low energy charge carriers of corrugated graphene have a similar behavior to electrons in strong gravitational fields \cite{PhysRevB.87.165131,AMORIM20161,VOZMEDIANO2010109}, providing a bridge between condensed matter and quantum gravity.    

The workhorse of straintronic theory is the Dirac equation in curved space-time, which gives a theoretical description of low energy charge carriers in deformed graphene \cite{Cortijo_2007,PhysRevB.76.165409,Vozmediano_2008,PhysRevB.82.073405,Gallerati2019}. This equation can be obtained as the low energy limit of the nearest neighbor tight-binding model with a space-dependent hopping parameter and nearest neighbor vectors. This space-dependence arises from the modification of the inter-atomic distance, which in turns, changes the hopping integrals. As long as these deformations are not too large, the ensuing metric and other differential geometry variables in the Dirac equation can be linked to the strain tensor and elasticity theory \cite{oliva,Naumis_2017}. 

Recently, the dynamics of electrons propagating in strained graphene samples has been studied numerically \cite{Flouris2018,Debus2018} and analytically \cite{Contreras_Astorga_2020}.  Remarkably, it was demonstrated that electrons can be confined by scattering on strained regions, allowing for electronic wave packet guiding. These investigations have been performed using simple strain field configurations having symmetries, simplifying the theoretical approaches. However, for controlling electrons in more complex applications, a framework for general surfaces is required. The main goal of this article is to provide theoretical strategies to study the dynamics of electrons in general strained surfaces using the Dirac equation in curved space-time.

Starting from the Dirac equation in curved space-time expressed in covariant notation, a comprehensive derivation of the Dirac equation in Cartesian coordinate for a general static graphene deformation is first presented. The non-trivial steps leading to an explicit equation are detailed, in particular when the system has no symmetry. Despite the apparent geometrical simplicity of such systems, consisting of a 2D arbitrary surface embedded in a 3D space, the resulting Dirac equation becomes quite intricate and challenging to solve analytically or numerically. Two strategies are introduced to simplify this problem: the diagonal metric approximation and the change of variables to isothermal coordinates. It is demonstrated that for both strategies, the Dirac equation has a simple form, reminiscent of the Dirac equation in flat space. However, they both have their own challenges: the diagonal approximation is valid only for a certain restricted class of surfaces while another equation, the Beltrami partial differential equation, needs to be solved to obtain the required quasi-conformal transformations, adding another layer of complexity. In this article, both approaches are described and analyzed. In addition, some specific examples and benchmark simulations are considered where both approaches are compared. The main outcome is a set of tools that can be applied for the computation of electron transport in deformed graphene samples.

This paper is organized as follows. Section \ref{sec:diff_geom} is devoted to a general presentation of the mathematical tools and framework required for the derivation of isothermal coordinates from a general 2D surface, including a definition of quasi-conformal transformations. In Section \ref{sec:dirac}, we propose a comprehensive derivation and analysis of the Dirac equation modeling graphene lattices in curved space.  A simplified version of the Dirac equation is also explicitly derived and justified when diagonal terms of the Riemannian metric tensor can be neglected. Section \ref{sec:algo} is devoted to the numerical approximation to the Beltrami and Dirac equations. The numerical approach is then tested and benchmarked in Section \ref{sec:numerics}, where the scattering of wave packets on local deformation is considered. We conclude in Section \ref{sec:conclusion}.

\section{Some elements of differential geometry for 2D surfaces}
\label{sec:diff_geom}

In this article, graphene is treated as a 2D curved surface embedded in a 3D Euclidean space. Therefore, before proceeding to the derivation of the Dirac equation, some important differential geometry results for surfaces are reviewed as they will be necessary in subsequent sections. 

Let us consider a 2D surface $\mathcal{S}$ embedded in a 3D Euclidean ambient space $\mathbb{R}^{3}$. Define a Cartesian coordinate chart $x,y,z$ on the ambient space. Then, the surface can be parametrized by
\begin{align}\label{surface}
\mathcal{S} &= \big\{(X(\boldsymbol{x}),Y(\boldsymbol{x}),Z(\boldsymbol{x})) \, \slash \, \boldsymbol{x} \in \mathcal{D} \big\} \, ,
\end{align}
with $X,Y,Z\in C^{1}(\mathcal{D};\R)$ and ${\boldsymbol x}=(x,y) \in \mathcal{D}\subset \mathbb{R}^{2}$. Denoting the vector in the ambient space ${\boldsymbol r}(\boldsymbol{x})=\big(X(\boldsymbol{x}),Y(\boldsymbol{x}),Z(\boldsymbol{x})\big) \in \mathbb{R}^{3}$, the Jacobian matrix of the transformation from the global to the local representation $J_{\boldsymbol r}^{\boldsymbol x}:=\cfrac{\partial {\boldsymbol r}}{\partial{\boldsymbol x}}$ reads
\begin{align}
	J_{\boldsymbol r}^{\boldsymbol x}  &= 
	\begin{bmatrix}
		X_x(\boldsymbol{x}) & Y_x(\boldsymbol{x}) & Z_x(\boldsymbol{x}) \\
		X_y(\boldsymbol{x}) & Y_y(\boldsymbol{x}) & Z_y(\boldsymbol{x})
	\end{bmatrix}
	,
\end{align}
where the notation $\Theta_{i}(\boldsymbol{x}) := \partial_{i} \Theta(\boldsymbol{x})$ (for $\Theta = X,Y,Z$ and $i=x,y$) has been introduced for simplicity.

Hence the naturally induced metric tensor $g_{\mathcal{S}}$ describing the surface locally is simply given by
\begin{align}
	g_{\mathcal{S}}(\boldsymbol{x}) &= J_{\boldsymbol r}^{\boldsymbol x}(J_{\boldsymbol r}^{\boldsymbol x})^T \\
	&=
	\begin{bmatrix}
		X_x^2+Y_x^2+Z_x^2  & X_xX_y +Y_xY_y+ Z_xZ_y\\
		X_xX_y +Y_xY_y + Z_xZ_y  & X_y^2+Y_y^2+Z_y^2 
	\end{bmatrix}
	\, .
\end{align}
To follow the traditional notation introduced by Gauss, the metric tensor in the frame $(\partial_x,\partial_y)$, is written as
\begin{align}
	\label{eq:metric_cart}
	g_{\mathcal{S}}(\boldsymbol{x}) &= 
	\begin{bmatrix}
		E(\boldsymbol{x})  & F(\boldsymbol{x})\\
		F(\boldsymbol{x}) & G(\boldsymbol{x}) 
	\end{bmatrix}
	\, ,
\end{align}
while the ensuing first fundamental form (metric tensor field) is given by
\begin{align}
\label{eq:general_diff_el}
ds^{2} = E(\boldsymbol{x}) dx^{2} + 2F(\boldsymbol{x}) dx dy + G(\boldsymbol{x})dy^{2}.
\end{align}
This metric describes locally a general surface in terms of the $\boldsymbol{x}$-coordinates, the original Cartesian coordinates of the plane $z=0$ in the ambient space. 
For general surface deformations, the non-diagonal term is non-zero ($F \neq 0$), implying that coordinates are not orthogonal over the whole domain. In addition, the presence of this non-diagonal term makes the calculations for the Dirac equation more tedious. In particular, the expression of Christoffel's symbols and the vielbein, required in the Dirac equation in curved space-time, has many terms and becomes complicated. For these reasons, it can be convenient to perform a change of variables to isothermal coordinates that diagonalizes the metric. 

\begin{definition}
	Let $\mathcal{S}$ be a 2D surface embedded in a 3D Euclidean ambient space. 
	Isothermal coordinates $\boldsymbol{u} = (u,v)$ are local orthogonal coordinates on $\mathcal{S}$ in which the metric is given by
	\begin{align}
	\label{eq:isothermal}
	ds^{2} = \rho(\boldsymbol{u}) \left[du^{2} + dv^{2}\right].
	\end{align}
\end{definition}
Using these coordinates entails the calculation of the function $\rho(\boldsymbol{u})$, as detailed below.
Nevertheless, because the metric is diagonal in this coordinate system, many equations such as the Dirac equation have a simpler form. Remarkably, it has been proven that for 2D surfaces, there always exists a (non-unique) coordinate change that allows for transforming the metric in the form of Eq. \eqref{eq:isothermal} \cite{ahlfors1955conformality,chern1955elementary} with a specific expression for $\rho$ in terms of the induced metric $g_{\mathcal{S}}$ in other coordinates (here, the Cartesian coordinates), the so-called quasi-conformal transformations  \cite{ahlfors2006lectures}. 
\begin{definition}
	Let the coordinates be expressed in the complex plane as $z=x+{\tt i}y \in \mathbb{C}$ and $w=u+{\tt i}v \in \mathbb{C}$. A mapping from the Cartesian to isothermal coordinates $z \rightarrow w$ is said quasi-conformal if it is a solution to the Beltrami equation: 
	\begin{align}\label{EB}
	w_{\bar{z}} = \mu(z) w_{z},
	\end{align}
	where $\Vert \mu\Vert_{\infty} < 1$ is the Beltrami coefficient.
\end{definition}
The solution to the Beltrami equation is a homeomorphism which preserves the orientation between Riemann surfaces with a {\it bounded} conformality distortion \cite{ahlfors2006lectures}.
%

The specific quasi-conformal transformation satisfied by isothermal coordinates, i.e. the expression for $\mu$, can be obtained explicitly. Introducing Wirtinger's derivatives
\begin{align}\label{wirtinger}
\partial_{z} = \frac{1}{2} \left[\partial_{x} - {\tt i} \partial_{y} \right], \qquad \partial_{\bar{z}} = \frac{1}{2} \left[\partial_{x} + {\tt i} \partial_{y} \right],
\end{align}
and using the chain rule, the metric \eqref{eq:general_diff_el} can be written as
\begin{align}
\label{eq:ds_gen_complex}
ds^{2} = \lambda |dz + \mu d\bar{z}|^{2},
\end{align} 
where
\begin{align}
\label{eq:beltrami_coeff}
\lambda = \frac{1}{4} \left(E + G + 2 \sqrt{\Delta}\right), \qquad \mu = \frac{E - G + 2{\tt i}F}{4\lambda},
\end{align}
and $\Delta = \mathrm{det}(g_{\mathcal{S}})$. On the other hand, the metric in isothermal coordinates \eqref{eq:isothermal} is given by
\begin{align}
ds^{2} &= \rho dw d\bar{w}, \\
\label{eq:ds_iso_complex}
&= \rho |w_{z}|^{2} \left| dz + \frac{w_{\bar{z}}}{w_{z}} d\bar{z}\right|^{2}.
\end{align}
Identifying Eqs. \eqref{eq:ds_gen_complex} and \eqref{eq:ds_iso_complex}, one can deduce that the mapping $z \rightarrow w$ produces isothermal coordinates as long as $\mu = w_{\bar{z}}/w_{z}$, i.e. as long as the mapping obeys the Beltrami equation. In addition, the comparison yields
\begin{align}\label{eq:isotherm_rho}
\rho(z) = \frac{\lambda(z)}{|w_{z}(z)|^{2}}.
\end{align}
Therefore, isothermal coordinates $w$ and their corresponding metric can be found explicitly by solving the Beltrami equation. 
%


\section{Dirac equation in curved space and strained graphene}\label{sec:dirac}

Applying an external force to a graphene sample produces a strain and deforms its atomic structure. The relative positions of carbon atoms in the lattice are modified, which in turns, changes the behavior of the  electrons travelling in the material. In a flat graphene sample, the low energy electrons are described in quantum mechanics by a massless Dirac equation, analogous to relativistic electrons \cite{RevModPhys.81.109}. Obviously, as graphene is subjected to mechanical constraints, one expects a different theoretical framework that takes strain into account. Remarkably, the quantum dynamical behavior of low energy electrons in a deformed graphene sheet is given by the Dirac equation in curved space \cite{Cortijo_2007,PhysRevB.76.165409,Vozmediano_2008,PhysRevB.82.073405,Gallerati2019,VOZMEDIANO2010109}. This important result has been obtained from the tight-binding model low energy limit \cite{oliva} and from general symmetry principles \cite{VOLOVIK2014352}. 

In this section, the Dirac equation describing strained graphene is stated, starting from the general covariant notation and specializing to some relevant cases for applications.


\subsection{Dirac equation in covariant notation}

In this section, the Dirac equation in curved space-time describing strained graphene is given in covariant notation \cite{pollock2010dirac}. Every tensor is expressed in terms of its components, with upper and lower indices denoting contravariant and covariant vector components, respectively. Einstein's summation convention on component indices is used hereafter. Three kinds of indices are used: Greek indices are general (local) curved space-time coordinates$/$charts ($\mu,\nu = 0,1,2$), uppercase latin indices are flat space Lorentz coordinates, while lowercase latin indices are the spatial coordinates of the general curved space ($i,j = 1,2$). In this formalism the metric ${\boldsymbol g}$ can be written ${\boldsymbol g}: T\mathcal{S} \times_{\mathcal{S}}T\mathcal{S}\rightarrow \R$, with ${\boldsymbol g}_q = g_{\mu\nu}(q)dx^{\mu}dx^{\nu}$, where $g_{\mu\nu}(q)={\boldsymbol g}_q(\partial_{\mu},\partial_{\nu})$ and ${\boldsymbol g}_{q}:T_{q}\mathcal{S}\times T_{q}\mathcal{S} \rightarrow \R$ with $q=(t,{\boldsymbol q})$ and ${\boldsymbol q}\in \mathcal{S}$.

In covariant notation, the curved space-time Dirac equation describing electrons in graphene in general curvilinear coordinates $q=(q^{1},q^{2},q^{3}) = (t,\boldsymbol{q})$ has a particularly simple form, which reads (in units where $\hbar = 1$)
\begin{align}
\label{eq:dirac_covariant}
{\tt i} \bar{\gamma}^{\mu}(q) D_{\mu} \psi(q) = 0,
\end{align}
where $\psi$ is the two-components wave function, $D_{\mu}$ is the covariant derivative (defined below) and $\bar{\gamma}^{\mu}(q) = (\bar{\gamma}^{0}(q),v_{F}\bar{\boldsymbol{\gamma}}(q))$, with $v_{F}$ the Fermi velocity, are the generalized gamma matrices. Introducing the metric tensor $g^{\mu \nu}(q)$ describing the surface locally, the generalized gamma matrices can be defined via their Clifford algebra \cite{tenseurs} as
\begin{align}
\label{eq:anticomm}
\{\bar{\gamma}^{\mu}(q), \bar{\gamma}^{\nu}(q) \} = 2 g^{\mu \nu}(q),
\end{align}
where $\{\cdot,\cdot \}$ is the anticommutator. Finding an explicit expression of these matrices can be performed in a local frame field by using the vielbein formalism. This permits a connection between generalized gamma matrices and flat space gamma matrices, given by
\begin{align}
\label{eq:gamma_vielbein}
\bar{\gamma}^{\mu}(q) = \gamma^{A}e_{ A}^{\mu}(q),
\end{align}  
where $e_{ A}(q)=e_{ A}^{ \mu}(q)\partial_{\mu}$ and $e_{ A}^{ \mu}=\partial x^{\mu}/\partial x^{A}$,  is the vielbein (spanning $T_q\mathcal{S}$ for any $q \in \mathcal{S}$). Similarly, orthonormal coordinates on the cotangent bundle $T^*_q\mathcal{S}$ are denoted $e^{A}(q)=e_{\mu}^{ A}(q)dx^{\mu}$. Some properties of the vielbein are summarized in Appendix \ref{app:prop_veil}.

The symbol $\gamma^{A}$ represents gamma matrices in flat space-time, obeying the usual relation 
\begin{align}
\{\gamma^{A}, \gamma^{B} \} = 2 \eta^{AB},
\end{align}
where $\eta = \mathrm{diag}(1,-1,-1)$ is the Minkowski metric for flat 2D space in Cartesian coordinates. An explicit representation of these matrices is given by the Dirac representation, where

\begin{align}
\gamma^{0} = 
\left(
\begin{array}{cc}
1 & 0 \\ 0 & -1  
\end{array}
\right)
\; \mbox{,} \;
\gamma^{1} = 
\left(
\begin{array}{cc}
0 & 1 \\ -1 & 0 
\end{array}
\right)
\; , \;
\gamma^{2} = 
\left(
\begin{array}{cc}
0 & -{\tt i} \\ -{\tt i} & 0
\end{array}
\right) \, .
\end{align}
This representation is used throughout this article.

One critical component of Eq. \eqref{eq:dirac_covariant} is the covariant derivative
\begin{align}
D_{\mu} = \partial_{\mu} + \Omega_{\mu}(q) - {\tt i}eA_{\mu}(q),
\end{align}
where  $\Omega_{\mu}(q)$ is the spinorial affine connection and $A_{\mu}$ is the four vector electromagnetic potential. The latter is included when an external electromagnetic field is coupled to electrons. On the other hand, the spinorial affine connection $\Omega_{\mu}(q)$ is introduced in the covariant derivative to preserve the covariance of the Dirac equation on curved space-time. In order to satisfy invariance by local Lorentzian transformations ($D_A\psi = L_A^BU(L)D_B\psi$, where $U(L)$ is the matrix representation of the Lorentz group), we get
\begin{align}
\label{eq:spin_connection}
\Omega_{\mu}(q) = -\frac{{\tt i}}{4} \omega_{\mu}^{\ AB}(q) \sigma_{AB},
\end{align}
where $\sigma_{AB} = {\tt i}[\gamma_{A},\gamma_{B}]/2$ is the commutator of the gamma matrices  while the spin connection is
\begin{align}
\omega_{\mu}^{\ AB}(q) =e_{\nu}^{A}(q) \left[   
\partial_{\mu}e^{\nu B}(q)  
+ \Gamma^{\nu}_{\ \mu \sigma}(q) e^{\sigma B}(q)
\right] \, ,
\end{align}
where the  Christoffel symbols 
\begin{align}
\label{eq:christo}
\Gamma^{\nu}_{\ \mu \sigma}(q) = \frac{g^{\nu \rho}(q)}{2} \left[
\partial_{\sigma}g_{\rho \mu}(q)
+ \partial_{\mu} g_{\rho \sigma}(q)
- \partial_{\rho}g_{\mu \sigma}(q)
\right] \, ,
\end{align}
were introduced.

This completes the description of the covariant Dirac equation in curved space-time. Each term of this equation can be evaluated from the metric and therefore, its explicit form depends on the surface deformation, and thus on the coordinate basis which is used.

\subsection{Static metric}

The Dirac equation \eqref{eq:dirac_covariant} applies to a general space-time and thus, includes effects from time-like deformation of the surface, when it is deformed dynamically. In this article, we restrict our analysis to static space-time curved surfaces. Then, the metric satisfies the conditions $\partial_{t} g_{\mu \nu}(q) = 0$, $g_{0i} = 0$ for $i=1,2$, and $g_{00} = 1$. In the following we denote by $g$ the static metric tensor (different from the general metric ${\boldsymbol g}$). It becomes
\begin{align}
\label{eq:full_metric}
g({\boldsymbol{q}}) =
\begin{bmatrix}
1 & 0 \\
0 & -g_{\mathcal{S}}(\boldsymbol{q})
\end{bmatrix},
\end{align} 
where the minus sign ensures that $g$ reduces to the Minkowski metric $\eta$ in the limit of zero curvature. 
For static metric of this form, the time-like gamma matrix is $\bar{\gamma}^{0}(q) = \gamma^{0}$ and the Dirac equation can be written in Schr\"odinger-like form, as
\begin{align}
\label{eq:dirac_H}
{\tt i}\partial_t \psi(t,{\boldsymbol q}) = H(t,{\boldsymbol q}) \psi(t,{\boldsymbol q}),
\end{align}
where $H(t,{\boldsymbol q})$ is the Dirac Hamiltonian operator in static curved space. It is defined as 
\begin{align}
H(t,{\boldsymbol q})  &=  
-{\tt i}v_{F}\bar{\alpha}^{i}(\boldsymbol{q})  \left[\partial_{i} + \Omega_{i}({\boldsymbol q}) - {\tt i} e A_{i}(t,{\boldsymbol q})\right] \nonumber \\
& -\mathbb{I}_{2} eA_{0}(t,{\boldsymbol q})   ,
\label{eq:dirac_hamil}
\end{align}
where the generalized Dirac matrices are (for $i=1,2$)
\begin{align}
\label{eq:gen_alpha}
\bar{\alpha}^{i}(\boldsymbol{q}) &:=  \gamma^{0}\bar{\gamma}^{i}(\boldsymbol{q}).
\end{align}
Eq. \eqref{eq:dirac_H} is the starting point of this article as it describes the quantum dynamics of electrons on curved static graphene surfaces. The deformed surface is characterized locally by the metric and thus, the expression of the covariant derivative and the generalized gamma matrices can be obtained from $g$.

\subsection{Dirac equation in Cartesian coordinates}\label{subsec:DECC}

In Cartesian coordinates $q = (t,\boldsymbol{x})=(t,x,y)$, according to Eqs. \eqref{eq:metric_cart} and \eqref{eq:full_metric}, the full space-time metric is 
\begin{align}
g(\boldsymbol{x}) = 
\begin{bmatrix}
1 & 0 &0 \\
0 & -E(\boldsymbol{x}) & -F(\boldsymbol{x}) \\
0 & -F(\boldsymbol{x}) & -G(\boldsymbol{x})
\end{bmatrix}.
\end{align}
To get an explicit expression of the Dirac equation, the vielbein has to be evaluated from Eq. \eqref{eq:vielbein}. Because the metric is not diagonal, this equation cannot be solved straightforwardly: first, it has to be written in matrix form and then diagonalized, as demonstrated in Appendix \ref{app:veil_cart}. 


%
The non-zero Christoffel symbols can be calculated from \eqref{eq:christo}, they are shown here for completion (where we have denoted $E_x$, $E_y$ etc, the partial derivatives of $E$ with respect to $x$, $y$):
\begin{align}\label{ChriSymb}
\Gamma^{1}_{11}(\boldsymbol{x}) &=
\frac{FE_{y} + GE_{x} - 2FF_{x}}{2\Delta},  \\
\Gamma^{2}_{12}(\boldsymbol{x}) = \Gamma^{2}_{21}(\boldsymbol{x}) &= 
\frac{EG_{x} - FE_{y}}{2\Delta} , \\
\Gamma^{1}_{22}(\boldsymbol{x}) &= 
\frac{2GF_{y} - FG_{y} - GG_{x}}{2\Delta},\\
\Gamma^{2}_{22}(\boldsymbol{x}) &= 
\frac{EG_{y}+FG_{x} - 2FF_{y}}{2\Delta},\\
\Gamma^{1}_{12}(\boldsymbol{x}) = \Gamma^{1}_{21}(\boldsymbol{x}) &=
\frac{GE_{y} - FG_{x}}{2\Delta} ,\\
\Gamma^{2}_{11}(\boldsymbol{x}) &=
\frac{2EF_{x} -EE_{y} - FE_{x}}{2\Delta} .
\end{align}
Finally, the last ingredient is the affine spin connection. Noticing that for a stationary (time-independent) surface, we have $g_{0i}=g^{0i}=0$ and $\Gamma^{0}_{ij} = \Gamma^{i}_{j0} = 0$, we hence obtain the following spin connection
\begin{align}
	\omega_1^{AB} &= e^{A}_{1}\partial_{1}e^{1B}+e^{A}_{2}\partial_{1}e^{2B}+e^{A}_1\Gamma_{11}^1e^{1B} \nonumber \\
	&+ e^{A}_2\Gamma_{11}^2e^{1B} + e^{A}_1\Gamma_{12}^1e^{2B}+ e^{A}_2\Gamma_{12}^2e^{2B} ,\\
	\omega_2^{AB} &= e^{A}_{1}\partial_{2}e^{1B}+e^{A}_{2}\partial_{2}e^{2B}+e^{A}_1\Gamma_{21}^1e^{1B} \nonumber \\
	&+ e^{A}_2\Gamma_{21}^2e^{1B} + e^{A}_1\Gamma_{22}^1e^{2B}+ e^{A}_2\Gamma_{22}^2e^{2B}  .
\end{align}
However, only the components $\omega_{1}^{12}$, $\omega_{2}^{12}$, $\omega_{1}^{21}$ and $\omega_{2}^{21}$ are required in the calculation of $\Omega_{i}$ because the Dirac matrix structure of the affine spin connection fulfills $\sigma_{11}= \sigma_{22} = 0$. As a matter of fact, the non-zero components are 
\begin{align}
\label{eq:dirac_struc}
	\sigma_{12} = \cfrac{i}{2}[\gamma_1,\gamma_2]  =  \gamma^{0} = -\sigma_{21}.
\end{align}
The expression of the vielbein and the spin connection can then be reported in Eqs. \eqref{eq:dirac_H} and \eqref{eq:dirac_hamil} to obtain the Dirac equation for a general static surface. Using results from the analysis of hyperbolic equations along with some assumptions on the regularity of the coefficients, one can show that the Cauchy problem of the resulting equation has a unique and regular solution (see Appendix \ref{app:reg_sol}). 

The Dirac equation in Cartesian coordinates is not shown here for simplicity, but can be evaluated using a computer algebra system.  However, going through this procedure is a tedious task and implementing the resulting Dirac equation numerically is very error-prone. For these reasons, two different strategies are now introduced to simplify this problem. In the first one, the non-diagonal terms in the metric are neglected while in the second one, a change of coordinates to isothermal coordinates (defined in Section \ref{sec:diff_geom}) is performed.

\subsection{Diagonal approximation in Cartesian coordinates}\label{subsec:diag}
In this section, non-diagonal terms $F$ of the metric tensor in Cartesian coordinates are neglected, allowing for a drastic simplification of the Dirac equation. The justification and conditions for this approximation are presented in Subsection \ref{sec:diag_approx}. 

\subsubsection{Dirac equation}
\label{sec:dirac_eq_diag}

Intuitively, the diagonal approximation can be performed when Cartesian coordinates are quasi-orthogonal everywhere on the surface, occurring when $|F | \ll \min ( | E |,| G |)$. In this case and under some continuity assumptions, it can be proven that the solution of the Dirac equation using the full metric is close to the solution obtained with the diagonal metric (see Section \ref{sec:diag_approx}). Then, the approximate metric simply becomes  
\begin{align}
\label{eq:metric_diag}
g(\boldsymbol{x}) = 
\begin{bmatrix}
1 & 0 &0 \\
0 & -E(\boldsymbol{x}) & 0 \\
0 & 0 & -G(\boldsymbol{x})
\end{bmatrix}.
\end{align}
For diagonal metrics, the expression of the vielbein can be easily determined from Eq. \eqref{eq:vielbein}. In this coordinate system, the natural local vielbein is also diagonal and is given by
\begin{align}
e(\boldsymbol{x}) =
\begin{bmatrix}
1 & 0 & 0 \\
0 & \sqrt{E(\boldsymbol{x})} & 0 \\
0 & 0 &  \sqrt{G(\boldsymbol{x})}
\end{bmatrix} .
\end{align}
Using these expressions of the metric and veilbein where it is assumed that non-diagonal terms of the metric $g$ on surface $\mathcal{S}$ \eqref{surface} are neglected, the Dirac equation modelling electrons on the graphene surface $\mathcal{S}$ reads
\begin{align}
\label{eq:dirac_cart_diag}
{\tt i}\partial_t \psi(t,{\boldsymbol x}) = \biggl\{
&-{\tt i}\frac{v_{F}}{\sqrt{E(\boldsymbol{x})}}\alpha^{1}   \left[\partial_{1} + \Omega_{1}({\boldsymbol x}) - {\tt i} e A_{1}(t,{\boldsymbol x})\right] \nonumber \\
&- {\tt i}\frac{v_{F}}{\sqrt{G(\boldsymbol{x})}}\alpha^{2}   \left[\partial_{2} + \Omega_{2}({\boldsymbol x}) - {\tt i} e A_{2}(t,{\boldsymbol x})\right] \nonumber \\
&-\mathbb{I}_{2}   eA_{0}(t,{\boldsymbol x})  
\biggr\} \psi(t,{\boldsymbol x}),
\end{align}
where the Dirac matrices in flat space are the Pauli matrices ($\alpha^{i} = \sigma^{i}$) and 
\begin{align}
\Omega_{1} &= \frac{{\tt i}}{4} \frac{E_{y}}{\sqrt{EG}} \gamma_{0}, \\
\Omega_{2} &= -\frac{{\tt i}}{4} \frac{G_{x}}{\sqrt{EG}} \gamma_{0}, 
\end{align}
and where $(A_0, A_i)$ represents an external electromagnetic field.

This form is similar to the Dirac equation in flat space, the only difference residing in the appearance of the $1/\sqrt{E}$ and $1/\sqrt{G}$ prefactors on the right-hand-side. This makes this approach very attractive from the computational point of view because the resulting equation can be straightforwardly evaluated from the surface. 


The expression of the Dirac equation is naturally much simpler when non-diagonal terms of the metric can be neglected. In the following subsection, we rigorously study the limits of the diagonal approximation of the metric tensor.  However, there exist surfaces where this approximation fails and the full metric, including non-diagonal terms, has to be taken into account. When this happens, it is convenient to perform a change of coordinates to isothermal coordinates, defined in Section \ref{sec:diff_geom}. 


\subsubsection{Diagonal approximation}
\label{sec:diag_approx}

Some analytical arguments are now given to justify the diagonal approximation in Cartesian coordinates described in Subsection \ref{sec:dirac_eq_diag}. Starting from Eq. \eqref{eq:dirac_H}, the full Dirac equation in curved space without approximation, for $(t,{\boldsymbol x}) \in [0;T]\times \mathcal{S}$, can be written as (neglecting the electromagnetic field)
\begin{align}\label{exact}
\partial_t \psi(t,{\boldsymbol x}) &  =  -  \biggl\{ \big(e_1^{1}({\boldsymbol x})\alpha^1 + e_2^{1}({\boldsymbol x})\alpha^2\big)\big(\partial_1+\Omega_1({\boldsymbol x})\big)  \nonumber   \\
&  \;\; \; +  \big(e_1^{2}({\boldsymbol x})\alpha^1 + e_2^{2}({\boldsymbol x})\alpha^2 \big)\big(\partial_2+\Omega_2({\boldsymbol x})\big) \biggr\}\psi(t,{\boldsymbol x}).
\end{align}
On the other hand, from the approximate diagonal metric tensor in Eq. \eqref{eq:metric_diag}, now written as $\widetilde{g}^{ij}_{\mathcal{S}}$, we obtain the following Dirac equation, for  $(t,{\boldsymbol x}) \in [0;T]\times \mathcal{S}$:
\begin{align}\label{approx}
\partial_t \widetilde{\psi}(t,{\boldsymbol x})  &=  -  \bigg\{\widetilde{e}_1^{1}({\boldsymbol x})\alpha^1\big(\partial_1+\widetilde{\Omega}_1({\boldsymbol x})\big)  \nonumber \\
&+ \widetilde{e}_2^{2}({\boldsymbol x})\alpha^2\big(\partial_2+\widetilde{\Omega}_2({\boldsymbol x})\big) \biggr\}\widetilde{\psi}(t,{\boldsymbol x}).
\end{align}
For $i \in \{1,2\}$, we define the following perturbation parameters $\vartheta, \epsilon$ by writing
\begin{align}\label{cond1}
\Omega_i({\boldsymbol x})  &=  \widetilde{\Omega}_i({\boldsymbol x}) + \vartheta_i({\boldsymbol x}), \\ e^{i}_A({\boldsymbol x})   &=  
\left\{
\begin{array}{lc} 
\widetilde{e}^{i}_A({\boldsymbol x}) + \varepsilon^{i}_A({\boldsymbol x}), & \hbox { if $i=A$}, \\
\varepsilon^{i}_A({\boldsymbol x}), & \hbox { if $i \neq A$},
\end{array}
\right.
\end{align}
where we assume that $\varepsilon$, $\vartheta$, $\widetilde{e}$ and $\widetilde{\Omega}$ belong to $L^2(\R^2,\C)\cap L^{\infty}(\R^2,\C^{2\times 2})$, and that there exists $\delta \in \mathbb{R}$ small enough, such that
\begin{align}\label{cond2}
\Vert \vartheta \Vert_{\infty} < \delta, \, & \, \Vert \varepsilon \Vert_{\infty} < \delta \, .
\end{align}
\begin{prop}
	We denote $\psi$ and $\widetilde{\psi}$ the respective solutions to \eqref{exact} and \eqref{approx} with the same smooth initial data. Then under the conditions \eqref{cond1}-\eqref{cond2},  there exists $C_T>0$ such that
	\begin{align}
		\sup_{t \in [0;T]}\| \psi(t,\cdot) - \widetilde{\psi}(t,\cdot)\|_{\infty} & \leq  C_T\delta \, .
	\end{align}
\end{prop}
{\bf Proof.} We set
\begin{align}
	\chi(t,{\boldsymbol x}) & :=   \psi(t,{\boldsymbol x}) - \widetilde{\psi}(t,{\boldsymbol x}) \, ,
\end{align}
and 
\begin{align}
	\phi(t,{\boldsymbol x}) &:=   \biggl\{ \big(\epsilon_1^{1}({\boldsymbol x})\alpha^1 + \epsilon_2^{1}({\boldsymbol x})\alpha^2\big)\big(\partial_1+\Omega_1({\boldsymbol x})\big)    \nonumber   \\
	&+  \big(\epsilon_1^{2}({\boldsymbol x})\alpha^1 + \epsilon_2^{2}({\boldsymbol x})\alpha^2 \big)\big(\partial_2+\Omega_2({\boldsymbol x})\big) \nonumber \\
	&+ \widetilde{e}_1^{1}({\boldsymbol x})\alpha^1\vartheta_1({\boldsymbol x})  + \widetilde{e}_2^{2}({\boldsymbol x})\alpha^2\vartheta_2({\boldsymbol x}) \biggr\}\psi(t,{\boldsymbol x}). 
\end{align}
Then, it is straightforward to show that $\chi$ satisfies
\begin{align}
\label{chi}
\begin{cases}
\partial_t\chi(t,{\boldsymbol x})   =  -\widetilde{e}_1^{1}({\boldsymbol x})\alpha^1\chi(t,{\boldsymbol x}) -\widetilde{e}_2^{2}({\boldsymbol x})\alpha^2\chi(t,{\boldsymbol x}) \\
\quad \quad \quad \quad \quad  - \phi(t,{\boldsymbol x}) \\
\chi(0,{\boldsymbol x})  =  {\boldsymbol 0} \, 
\end{cases},
\end{align}
where $\phi \propto \varepsilon, \vartheta$ contains all the terms proportional to the perturbation parameters.
Next, because the solution $\psi$ to the Dirac equation is in $H^1(\R_+\times \R^2,\C^2)$, there exists $C_1>0$, such that $\|\phi\|_{1}\leq C_1\delta$, where $\|\, \cdot \, \|_1$ is the $H^1$-norm. We simply conclude using Gronwall's lemma on the characteristic surface for \eqref{chi}, which then allows us to deduce that $\|\chi\|_1 \leq C_2\delta$ for some $C_2>0$.  $\Box$

These arguments based on perturbation theory allow us to conclude that the diagonal approximation of the metric is relevant whenever $|||g^{ij}_{\mathcal{S}}-\widetilde{g}^{ij}_{\mathcal{S}}|||$ is small enough, occurring when $|F| \ll |E|$ and  $|F| \ll |G|$.

\subsection{Dirac equation in isothermal coordinates}

Isothermal coordinates have been defined in Section \ref{sec:diff_geom} as coordinates where the metric is diagonal. A mapping from any coordinate system to these coordinates always exists for 2D surface embedded in a 3D Euclidean space and can be obtained from a solution of the Beltrami equation \eqref{EB}. Therefore, starting from Cartesian coordinates for which the surface parametrization is more natural, we switch to isothermal coordinates, where the Dirac equation has a simpler form. The latter is now given explicitly.

In isothermal coordinates $q = (t,\boldsymbol{u})=(t,u,v)$, the spatial part of the line element is given by \eqref{eq:isothermal}. Therefore, the corresponding metric is diagonal and its matrix representation is
\begin{align}
g(\boldsymbol{u}) = 
\begin{bmatrix}
1 & 0 &0 \\
0 & -\rho(\boldsymbol{u}) &0 \\
0 & 0 & -\rho(\boldsymbol{u})
\end{bmatrix}.
\end{align}
Just like in the diagonal approximation, an expression of the natural vielbein can be easily found:
\begin{align}
e(\boldsymbol{u}) =
\begin{bmatrix}
1 & 0 & 0 \\
0 & \sqrt{\rho(\boldsymbol{u})} & 0 \\
0 & 0 &  \sqrt{\rho(\boldsymbol{u})}
\end{bmatrix} .
\end{align}
Because both the metric and vielbein are diagonal in isothermal coordinates, the Dirac equation can then be simplified to
\begin{align}
\label{eq:dirac_iso}
{\tt i}\partial_t \psi(t,{\boldsymbol u}) &= \biggl\{
-{\tt i}\frac{v_{F}}{\sqrt{\rho(\boldsymbol{u})}}\alpha^{i}   \left[\partial_{i} + \Omega_{i}({\boldsymbol u}) - {\tt i} e A_{i}(t,{\boldsymbol u})\right] \nonumber \\
&-\mathbb{I}_{2}   eA_{0}(t,{\boldsymbol u})  
\biggr\} \psi(t,{\boldsymbol u}) \, ,
\end{align}
where the affine spin connection is
\begin{align}\label{Omegai}
\Omega_{1}(\boldsymbol{u}) &= \frac{{\tt i}}{4} \frac{\rho_{v}(\boldsymbol{u})}{|\rho(\boldsymbol{u})|} \gamma_{0}, \\
\Omega_{2}(\boldsymbol{u}) &= -\frac{{\tt i}}{4} \frac{\rho_{u}(\boldsymbol{u})}{|\rho(\boldsymbol{u})|} \gamma_{0}.
\end{align}
Again, this equation in curved space is similar to the Dirac equation in flat space, making isothermal coordinates very attractive from the computational point of view as they minimize the number of terms in the equation.

The formulation in isothermal coordinates can be applied to general surfaces and leads to a greatly simplified expression of Christoffel's symbols and spin connections, similar to the cartesian case when the non-diagonal terms of the metric are neglected. The main challenge however lies in the solution to the Beltrami equation, required  to construct the mapping $\boldsymbol{x} \rightarrow \boldsymbol{u}$ and the function $\rho({\boldsymbol u})$. There exists a few analytical solutions to the Beltrami equation for simple configurations, but for general surfaces, it has to be obtained numerically. A numerical scheme for solving the Beltrami equation is presented in Section \ref{sec:num_beltrami}.

\section{Numerical schemes}\label{sec:algo}

It is very challenging to solve analytically the time-dependent Dirac equation in curved space-time. The existing solutions are for mostly for highly symmetric and static systems. Therefore, to study the electronic dynamics in general configuration, we now resort to an accurate numerical approach. 

\subsection{Numerical method for the Dirac equation}

In this section, a numerical scheme is presented to solve the Dirac equation for strained graphene given in Eqs. \eqref{eq:dirac_cart_diag} and \eqref{eq:dirac_iso}. To reach this goal, it is convenient to consider the corresponding Dirac Hamiltonian:
\begin{align}
H &= -{\tt i}\frac{v_{F}}{\sqrt{E(\boldsymbol{x})}}\alpha^{1}   \left[\partial_{1} + \Omega_{1}({\boldsymbol x}) - {\tt i} e A_{1}(t,{\boldsymbol x})\right] \nonumber \\
&- {\tt i}\frac{v_{F}}{\sqrt{G(\boldsymbol{x})}}\alpha^{2}   \left[\partial_{2} + \Omega_{2}({\boldsymbol x}) - {\tt i} e A_{2}(t,{\boldsymbol x})\right] \nonumber \\
&-\mathbb{I}_{2}   eA_{0}(t,{\boldsymbol x}) . 
\end{align}
Strictly speaking, this Hamiltonian is not Hermitian ($H^{\dagger} \neq H$), stemming from the fact that the self-conjugation of the affine spin connection is not Hermitian: it can be demonstrated that $\Omega_{i}^{\dagger} = -\gamma^{0} \Omega_{i} \gamma^{0}$, a property that breaks the hermiticity of the Hamiltonian operator. Nevertheless, it is known that $H$ is self-adjoint ($\langle \phi |H^{\dagger} | \psi \rangle_{g} = \langle \phi |H | \psi \rangle_{g}$) with respect to the $g$-scalar product \cite{PhysRevD.22.1922}
\begin{align}
\langle \phi |\psi \rangle_{g} &= \int dx \sqrt{\mathrm{det}(g)} \phi^{\dagger}(x) \psi(x).
\end{align}
Of course, one recovers the $L^{2}$ product in flat space when $\mathrm{det}(g) = 1$. The corresponding $g$-norm $\Vert \psi \Vert_{g} = \langle \psi |\psi \rangle_{g}$ is conserved by the time-evolution given by the Dirac equation in curved space \cite{PhysRevD.22.1922}. However, it is not straightforward to develop a numerical scheme that explicitly conserves the $g$-norm because the Hamiltonian is not Hermitian. As a consequence, the discrete time evolution operators in usual strategies like operator splitting or the Crank-Nicolson method, will not be unitary and thus, will not conserve the norm. 

This problem can be cured by noticing that the Hamiltonian is a pseudo-Hermitian operator \cite{PhysRevD.83.105002}. Such operators can be transformed as
\begin{align}
H_{\eta} = \eta H \eta^{-1},
\end{align} 
such that $H_{\eta}^{\dagger} = H_{\eta}$. This $\eta$-representation incurs a transformation of the norm, which becomes the usual $L^{2}$-norm while the wave function becomes $\psi_{\eta} = \eta \psi$. Then, it can be demonstrated that the scalar products are identical:
\begin{align}
\langle \phi | \psi \rangle_{g} = \langle \phi_{\eta} | \psi_{\eta} \rangle_{L^{2}},
\end{align}
as long as the time evolution is performed by $H$ for $\phi,\psi$ and by $H_{\eta}$ for $\phi_{\eta},\psi_{\eta}$. Given that $H_{\eta}$ is Hermitian, usual techniques can thus be used and should preserve the norm explicitly. 

The transformation that Hermiticizes the Hamiltonian is $\eta = (\mathrm{det}g)^{\frac{1}{4}}$ \cite{PhysRevD.83.105002}, yielding
\begin{align}
H_{\eta} &= (\mathrm{det}g)^{\frac{1}{4}} H (\mathrm{det}g)^{-\frac{1}{4}} \\
&= -{\tt i}\frac{v_{F}}{\sqrt{E(\boldsymbol{x})}}\alpha^{1}   \left[\partial_{1}
- \frac{\partial_{1} \sqrt{\mathrm{det}g}}{2 \sqrt{\mathrm{det}g} }
+ \Omega_{1}({\boldsymbol x}) - {\tt i} e A_{1}(t,{\boldsymbol x})\right] \nonumber \\
&- {\tt i}\frac{v_{F}}{\sqrt{G(\boldsymbol{x})}}\alpha^{2}   \left[\partial_{2}
- \frac{\partial_{2} \sqrt{\mathrm{det}g}}{2 \sqrt{\mathrm{det}g} }
+ \Omega_{2}({\boldsymbol x}) - {\tt i} e A_{2}(t,{\boldsymbol x})\right] \nonumber \\
&-\mathbb{I}_{2}   eA_{0}(t,{\boldsymbol x}). 
\end{align}
The numerical scheme will thus solve the equation
\begin{align}
\label{eq:dirac_eta}
i \partial_{t} \psi_{\eta} = H_{\eta} \psi_{\eta}.
\end{align}
For this purpose, the Pseudo-spectral Crank-Nicolson (PSCN) method introduced in Ref. \cite{jcp2019,jcp2020} is used. Introducing a set of discrete times given by $t_{n} = t_{0} + n\Delta t$ for $n \in \mathbb{Z}^{+}$, where $t_{0}$ is the initial time and $\Delta t$ is the time step, the Crank-Nicolson update is written as
\begin{align}
\label{eq:cn}
\left(1 + \mathtt{i}\frac{\Delta t}{2} H_{\eta} \right) \psi_{\eta}^{n+1} = \left(1 - \mathtt{i}\frac{\Delta t}{2} H_{\eta} \right) \psi_{\eta}^{n},
\end{align}
where $\psi_{\eta}^{n} = \psi_{\eta}(t_{n})$. 

The space discretization is performed by introducing a 2D grid where each point is given by $(x_{k},y_{l}) = (x_{0} + k \Delta x, y_{0} + l \Delta y)$ for $k,l \in \mathbb{Z}^{+}$, where $x_{0},y_{0}$ are the position of the boundary and $\Delta x, \Delta y$ are the spatial steps. The wave function is projected on the grid, $\psi^{n}_{\eta,kl} = \psi(t_{n},x_{k},y_{l})$, yielding a linear system of the form
\begin{align}
L \boldsymbol{\psi}^{n+1} = \boldsymbol{b},
\end{align}  
whose solution gives the updated wave function $\psi_{\eta, kl}^{n+1}$.

The gist of the PSCN method is the construction of the matrix $L$ and the vector $\boldsymbol{b}$ by using a spectral method to evaluate spatial derivatives. In particular, the derivative are projected on the grid via 
\begin{equation}
[\partial_{i} \psi^{n}]_{kl} = \mathcal{F}_{i}^{-1}\left[\mathtt{i}k_{i} \mathcal{F}_{i}\left[ \psi^{n} \right] \right]_{kl},
\end{equation}
where $k_{i}$ are the discrete Fourier modes in axis $i$ while $\mathcal{F}_{i}$ is the partial discrete Fourier transform operator that performs a Fourier thansform in the $i$-direction. Armed with this notation, the right-hand-side of Eq. \eqref{eq:cn} can be evaluated straightforwardly, assuming that $\psi^{n}_{\eta,kl}$ is known. The left-hand-side is more challenging: the solution is defined implicitly and therefore, the operation $[\partial_{i} \psi^{n+1}]_{kl}$ is not known {\it a  priori}. The strategy introduced in \cite{jcp2020} consists of using a Krylov iterative technique, such as GMRES or Conjugate Gradient, to solve the linear system. This is interesting for two main reasons:
\begin{itemize}
	\item In these methods, an initial guess is chosen and improved iteratively towards the solution. Here, the initial guess is chosen as $\psi^{n}$ which is close to $\psi^{n+1}$ for small $\Delta t$ and on which the discrete spectral derivative can be evaluated.
	
	\item These methods allow for matrix-free solution of the linear system, where the matrix $L$ is not constructed explicitly nor stored in memory. Rather, a linear operator is defined which yield the vector $\boldsymbol{v} = L \boldsymbol{\psi}^{n,m}$, where $\boldsymbol{\psi}^{n,m}$ is the $m$-th iteration of the Krylov method. Most implementation of Krylov methods allows for defining such operators. 
\end{itemize}
The PSCN scheme has second order convergence in time and spectral convergence in space. In addition, when the fast Fourier transform (FFT) is used to compute the derivatives, the complexity per time step is $O(N \ln N)$, where $N$ is the total number of grid points. The numerical method was implemented in Python using the FFTW interface\footnote{{\tt www.fftw.org}} for the FFT and the matrix-free GMRES algorithm was used to solve the linear system. We again refer the reader to \cite{jcp2020} for more details and analysis of the method.

\subsection{Numerical approximation of the Beltrami equation} \label{sec:num_beltrami}

In Sections \ref{sec:diff_geom} and \ref{sec:dirac}, it was argued that using isothermal coordinates requires a solution to the Beltrami equation. The latter gives the mapping between Cartesian coordinates, where a natural parametrization of the surface exists, and the isothermal coordinates, where the Dirac equation has a simple form. However, explicit solutions to the Beltrami equation are usually challenging to find due to the mathematical complexity of the system of equations. As a consequence, those calculations need to be performed numerically. In this section, a numerical scheme to perform this task is presented.

The overall strategy$/$algorithm is now summarized:
\begin{enumerate}
	\item Parametrization of the surface $\mathcal{S}$ in Cartesian coordinates, then explicit construction of $E$, $F$, $G$, and $g$.
	\item Evaluation of the Beltrami coefficient $\mu$ via Eq. \eqref{eq:beltrami_coeff}.
	\item Numerical computation of $w_h$, an approximate solution to the Beltrami equation. 
	\item Computation of $\rho_h$ via Eq. \eqref{eq:isotherm_rho}.
	\item Evaluation of $\rho_h$ as a function of $u$ and $v$, which requires the inverse of $w=u+{\tt i}v$.
	\item Estimation of the Dirac equation in isothermal coordinates thanks to a then diagonal metric tensor.
\end{enumerate}
Each step is relatively straightforward, except the third one, which is now detailed.

\subsubsection{Numerical solution to the Beltrami equation}

A convergent numerical scheme to solve the Beltrami equation \eqref{EB} numerically is now discussed. It is assumed that the complex dilation $\mu$ given in \eqref{EB}  is such that $\mu\in L^{\infty}(\C)$ with $\|\mu\|_{\infty}  <1$. According to \cite{ahlfors1955conformality,ahlfors2006lectures}, there exists a unique quasiconformal mapping $w$  satisfying the Beltrami equation with values fixed at $z=0$, $1$ and at infinity.  For $\mu$ analytic, one of the most standard, simple and efficient methods is the one derived by Daripa \cite{daripa,Gaidashev}, based on numerical approximation of Hilbert's and Cauchy's transforms. However, the downside of this approach are the required strong regularity of $\mu$ as well as the need for working on unbounded domains. Going beyond these restrictions, the most natural alternative is the least-square finite-element method based on a variational formulation of the Beltrami equation.

To obtain this variational formulation, the real and imaginary parts of the Beltrami equation are separated and formulated as a system of partial differential equations. Assuming that $|\mu| < 1$ over $\Omega\subset \C$, the Beltrami equation \eqref{EB} is rewritten as \cite{origami}
\begin{align}
	\label{eq:bel_sys}
	\nabla u(\boldsymbol{x}) &= JA \nabla v(\boldsymbol{x}),
\end{align}
where $A$ and $J$ are defined as
\begin{align}
	A  &= 
	\cfrac{1}{1-|\mu|^2} \nonumber \\
	&\times 
	\begin{bmatrix}
		\bigl(\mathrm{Re}(\mu)-1 \bigr)^2 + \mathrm{Im}(\mu)^2 & -2\mathrm{Im}(\mu) \\
		-2\mathrm{Im}(\mu) & \bigl(1+\mathrm{Re}(\mu) \bigr)^2 +\mathrm{Im}(\mu)^2
	\end{bmatrix}
	,  \\
	J &= 
	\begin{bmatrix}
		0 & -1 \\
		1 & 0 
	\end{bmatrix}
	\, .
\end{align}
Recall that there is no unique choice of isothermal coordinates as they depend on the choice of boundary values. The Beltrami equation is solved in Cartesian coordinates on a rectangular domain given by $\Omega = [x_{\mathrm{min}},x_{\mathrm{max}}] \times [y_{\mathrm{min}},y_{\mathrm{max}}]$.  
To obtain a unique solution, the following mixed Dirichlet and Neumann boundary conditions are considered: 
\begin{align}
\label{eq:bel_bound_u}
\left. u \right|_{\partial \Omega^{\mathrm{l}}} &= x_{\mathrm{min}}, & 
\left. u \right|_{\partial \Omega^{\mathrm{r}}} &= x_{\mathrm{max}} ,  \nonumber \\ 
\left. u_{y} \right|_{\partial \Omega^{\mathrm{d}}} &= 0, &
\left. u_{y} \right|_{\partial \Omega^{\mathrm{t}}} &= 0 ,& \\
\label{eq:bel_bound_v}
\left. v_{x} \right|_{\partial \Omega^{\mathrm{l}}} &= 0, &
\left. v_{x} \right|_{\partial \Omega^{\mathrm{r}}} &= 0 , \nonumber \\ 
\left. v \right|_{\partial \Omega^{\mathrm{d}}} &= y_{\mathrm{min}},& 
\left. v \right|_{\partial \Omega^{\mathrm{t}}} &= y_{\mathrm{max}} ,&
\end{align}
where $\partial \Omega^{\mathrm{l,r,d,t}}$ are the left, right, down and top boundaries of the rectangular domain, respectively. These boundary conditions allow for mapping the rectangular domain $\Omega$ in Cartesian coordinate to the same domain in isometric coordinates ($\Omega \rightarrow \Omega$), such that $\boldsymbol{u} \in \Omega$ in isothermal coordinates. It also ensures that for a flat metric, we have $x=u$ and $y=v$. Finally, it assumes that the deformation has a support such that $\mu |_{\partial \Omega} = 0$. These boundary conditions can be written in the form $\mathcal{R}\boldsymbol{u} = \boldsymbol{h} $ on $\partial \Omega$, where $\mathcal{R}$ is a differential operator and $\boldsymbol{h}$ is a set of constants. Similar boundary conditions have been considered in Refs. \cite{mastin1978discrete,doi:10.1137/120866129}. 

This equation can hence be solved using different approaches. In this work, a standard least square finite element method is considered, which is well-suited for PDE with first order differentials \cite{bochev2006least}. Using the properties of $A$ and $J$, the least square functional is written as 
\begin{align}\label{Ltilde}
	\mathcal{L}(u,v;\mu) &= \Vert P\nabla u+JP\nabla v \Vert_{L^{2}_{\Omega}}^2 + \Vert \mathcal{R}\boldsymbol{u} - \boldsymbol{h} \Vert^{2}_{L^{2}_{\partial \Omega}} ,
\end{align}
where:

\begin{enumerate}
	\item The function $(u,v)$ belongs to $V$, defined as
\begin{align}
		V=\big\{w \in H^1(\Omega;\R)  \big\}\times \big\{w \in H^1(\Omega;\R)\big\} \, .
\end{align}
\item The norm $\| \cdot \|_{L^{2}_{\Omega}}$ is a norm on $\big(\L^2(\Omega;\C)\big)^2$.
\item The matrix $P$ satisfies $P^TP=A$ and is explicitly given by \cite{origami}
\begin{align}
	P = 
	\cfrac{1}{\sqrt{1-|\mu|^2}}
	\begin{bmatrix}
		1-\mathrm{Re}(\mu) & -\mathrm{Im}(\mu) \\
		-\mathrm{Im}(\mu) & 1+\mathrm{Re}(\mu)
	\end{bmatrix}
	\, .
\end{align}
\end{enumerate}
The functional $\mathcal{L}$ is minimized when $(u,v) \in V$ is solution to Eq. \eqref{EB} with boundary conditions \eqref{eq:bel_bound_u}-\eqref{eq:bel_bound_v}. Following \cite{bochev2006least}, the corresponding Euler-Lagrange equation is
\begin{align}
\label{eq:func_bel}
\bigl(P\nabla \widehat{u}+JP\nabla \widehat{v} , P\nabla u+JP\nabla v\bigr)_{L^{2}_{\Omega}} \nonumber \\
+
\bigl(\mathcal{R}  \widehat{\boldsymbol{u}} , 
\mathcal{R}  \boldsymbol{u}\bigr)_{L^{2}_{\partial \Omega}}
&= \bigl(\mathcal{R}  \widehat{\boldsymbol{u}} , \boldsymbol{h}\bigr)_{L^{2}_{\partial \Omega}} ,
\end{align}
where $\widehat{u},\widehat{v}$ are test functions in $H^1(\Omega;\R)$. Then, the discretization proceeds as usual for finite element methods:
\begin{enumerate}
	\item A triangulation of $\Omega$ is introduced.
	\item The functions $u, v$ are expanded on a piecewise polynomial basis.
	\item A gradient descent-like method is used to solve the resulting linear system of equations. 
\end{enumerate}
In this work, a finite element method and piecewise continuous second order polynomials is chosen as basis functions. More specifically, $u$ and $v$ are approximated by piecewise continuous quadratic polynomials ($P_2-P_2$):
\begin{align}
u_h &= \sum_{i\in \tau_{I}} c_i u_i ,\\
v_h &= \sum_{i\in \tau_{I}}d_i v_i,
\end{align}
where $\{c_{i}\}_, \{d_{i}\}_i$ are expansion coefficients and $\{(u_{i},v_{i})\}_i \in P_{2}$ are the polynomial basis. This is performed on a conformal triangular finite element mesh $\tau_h=\tau_h(\Omega)$ whose elements are indexed by a finite set $I=\{1,\cdots,\textrm{dim}(V_h)\}$, and such that $V_h \subset V$. In order to minimize a finite dimensional version of $\min_{(u,v)\in V} \mathcal{L}(u,v;\mu)$, we apply a least square method on $(u_h,v_h)\in V_h$, that is we minimize
\begin{align}
	\min_{(u_h,v_h) \in V_h} \left( \Vert P_h\nabla u_h + JP_hv_h \Vert^2_{L^{2}_{\tau_{h}}}  + \Vert \mathcal{R} \boldsymbol{u}_{h} - \boldsymbol{h}_{h} \Vert^{2}_{L^{2}_{\partial \tau_{h}}} \right)
\end{align}
where $\Vert \cdot \Vert^{2}_{L^{2}_{\tau_{h}}}$ denotes the $L^2$-norm on $\tau_h$. This leads to finding a non-trivial solution to ${\bf L}{\bf x}={\bf f}$ with matrix ${\bf L}=\{L_{ij}\}_{ij}$ and vectors ${\bf f} = \{f_{i}\}_{i}$ constructed from \eqref{eq:func_bel}, that is for 
\begin{align}
	L_{ij} &= \big(P\nabla \widehat{u}_i + JP\nabla \widehat{v}_i\, ,\, P\nabla u_j + JP\nabla v_j \big)_{L^2_{\tau_{h}}} \nonumber \\
	&+
		\big(\mathcal{R} \widehat{\boldsymbol{u}}_i  , \mathcal{R} \widehat{\boldsymbol{u}}_j \big)_{L^2_{\partial \tau_{h}}} ,\\
	f_{i} &= \big(\mathcal{R} \widehat{\boldsymbol{u}}_i  , \boldsymbol{h} \big)_{L^2_{\partial \tau_{h}}} ,
\end{align}
for basis functions $(u_i,v_i)$ of $V_h$ and $(\widehat{u}_i,\widehat{v}_i)$ of $V_h$ dense in $\big(H^1(\Omega;\R)\big)^2$. 

Our implementation of this numerical method is based on the finite element package F\textsc{enics} \cite{AlnaesBlechta2015a}, which has a simple interface allowing for a symbolic definition of the functional \eqref{eq:func_bel}, along with the possibility of using many element types and discretization. In particular, the mesh and the linear system are generated automatically from the specification of the domain and the functional. Also, the resulting code is parallelized using the message passing interface (MPI), allowing for good performance on large problems. Finally, the linear system is solved using a Krylov iterative method (GMRES).

\subsection{Convergence of the Beltrami and Dirac solvers}

The analysis of convergence of the Beltrami equation solver is standard, we hence only provide the main results. Let us first notice that the operator $\mathcal{L}$ \eqref{Ltilde}, is continuous and coercive with respect to $u$ and $v$ in $H^1(\Omega)$ such that for some constants $c(\mu)>0$ and $\alpha(\mu;\Omega)>0$
\begin{align}
|\mathcal{L}(u,v;\mu)| & \leq  c(\mu) \|u\|_{H^1(\Omega)}\|v\|_{H^1(\Omega)} \\
|\mathcal{L}(u,v;\mu)| & \geq  \alpha(\mu;\Omega) (\|u\|_{H^1(\Omega)} + \|v\|_{H^1(\Omega)}) \, . 
\end{align}
Continuity is a consequence of the fact that $\mu$ belongs to $L^{\infty}(\Omega)$, while coercivity comes from Poincar\'e's inequality on $\Omega$ bounded \cite{brezis}. By Lax-Milgram's theorem we deduce the existence of a unique solution to the Beltrami equation in $H^1(\Omega)\times H^1(\Omega)$, \cite{raviart}. Assuming that $u$ and $v$ belong to $H^{s+1}(\Omega)$, the least square finite element solution $(u_h,v_h)$ is such that (see \cite{LeastSquare})
\begin{align}
\|u-u_h\|_{L^2(\Omega)} + \|v-v_h\|_{L^2(\Omega)}  \leq \nonumber \\  c(\mu;\Omega)h^{s+1}(\|u\|_{H^{s+1}(\Omega)} + \|v\|_{H^{s+1}(\Omega)}) \, ,
\end{align}
and
\begin{align}
\|u-u_h\|_{H^1(\Omega)} + \|v-v_h\|_{H^1(\Omega)}  \leq \nonumber \\
c(\mu;\Omega)h^{s}(\|u\|_{H^{s+1}(\Omega)} + \|v\|_{H^{s+1}(\Omega)}) \, .
\end{align}
In numerical experiments presented in the next section, we are using $P^2$ finite elements, and we indeed observe a third order convergence.

Regarding the Dirac equation solver, let us recall some basic facts about the pseudo-spectral method  used in this paper. First, because $H_{\eta}$ is hermitian, the $\ell^2$-norm of the numerical solution $\psi_h$ is trivially conserved using a trapezoidal rule: $\|\psi_h^{n}\|_{\ell^2} = \|\psi_h^{0}\|_{\ell^2}$.

Regarding pseudo-spectral methods, let us recall that for $\psi$ smooth enough and a pseudo-spectral approximation on a $N$-point grid $\psi_h$, we have \cite{bardos}
\begin{align}
\|\psi_h-\psi\| & \leq & N^{s-r}\|\psi\|_{H^{s}} \, , \, \textrm{ for } s>r>d/2 \in \R \, .
\end{align}
We do not provide an analysis of the convergence of the Dirac equation solver, as it would require very important technical effort and is outside the scope of this article. However, we recall a standard result on first order one-dimensional linear hyperbolic equations, which provides interesting information regarding the accuracy of the overall pseudo-spectral approach used in our paper. Let us recall that the studied Dirac equation is a linear hermitian hyperbolic system but \textit{a priori} non-conservative. It was proven in \cite{GHT94,bardos}, that for $\partial_t v + \partial_x (q(x)v) = 0$ with $v(0,\cdot)=v_0$, a pseudo-spectral method in space leads to the following error estimate
\begin{align}
\|v_h(t,\cdot)-v(t,\cdot)\|_{L^2}  \leq \nonumber \\
e^{\|q'\|_{\infty}t}\|\big(N^{1-s}\|v_0\|_{H^{s}}+ N^{2-s}\max_{\tau \leq t}\|v(\tau,\cdot)\|_{H^s}\big)  \, .
\end{align}
Interestingly, it is shown in \cite{bardos} that the pseudo-spectral scheme looses one order of convergence compare to a full spectral method. Overall, we expect the same typical convergence accuracy in space for the Dirac equation solver, as long as their coefficients are smooth and bounded. In time, as we use a trapezoidal rule, we trivially have a second order convergence. 

We then conjecture that the error on the pseudo-spectral approximation of $\psi$ is bounded by (for $s > 1 + d/2$):
\begin{align}
\|\psi_h(t,\cdot)-\psi(t,\cdot)\|^2_{L^2}  \leq  \nonumber \\
e^{2Q_{\infty}t}\|\big(N^{-2s}\|\psi_0\|_{H^{s}}+ N^{1-s+d/2}\max_{\tau \leq t}\|\psi(\tau,\cdot)\|_{(H^s)^4}\big) \, ,
\end{align}
where 
\begin{align}
Q_{\infty} := \max\big\{ |||\nabla A |||_{\infty},|||\nabla B |||_{\infty},|||\nabla C |||_{\infty} \big\} \, .
\end{align}

\section{Numerical results}\label{sec:numerics}

In this section, some numerical experiments are presented to test the numerical approaches presented earlier. 

\subsection{Isothermal coordinates}

The first numerical results are focused on the solution of the Beltrami equation. We consider two different numerical tests in which the convergence of the solver is assessed empirically.

\subsubsection{Convergence of the Beltrami solver}

To verify the numerical method and the implementation of the Beltrami equation least-square finite element solver, a simple test case is introduced. This is performed via the method of manufactured solution \cite{salari2000code}, where an explicit solution is selected while the parameters in the partial differential equation are fixed from this solution. For the Beltrami equation given in Eq. \eqref{EB}, the particular solution considered in this article is
\begin{align}
u_{p}(x,y) &= x - x_{0} + C_{u} \sin^{2}\left(\frac{\pi x}{L_{x}}\right)\sin^{2}\left(\frac{\pi y}{L_{y}}\right), \\
v_{p}(x,y) &= y - y_{0} + C_{v} \sin^{2}\left(\frac{\pi x}{L_{x}}\right)\sin^{2}\left(\frac{\pi y}{L_{y}}\right), 
\end{align} 
where $C_{u},C_{v} \lesssim 1$ are arbitrary constants, $x_{0} = x_{\mathrm{min}} +  L_{x}/2$ and $y_{0} = y_{\mathrm{min}} + L_{y}/2$, where we defined $L_{x} = |x_{\mathrm{max}} - x_{\mathrm{min}}|$ and $L_{y} = |y_{\mathrm{max}} - y_{\mathrm{min}}|$. This particular solution obeys the boundary conditions in Eqs. \eqref{eq:bel_bound_u}-\eqref{eq:bel_bound_v}. The Beltrami coefficient is obtained from the solution via
\begin{align}
\mu_{p}(x,y) = \frac{u_{p,x} - v_{p,y} + \mathtt{i}(v_{p,x} + u_{p,y})}{u_{p,x} + v_{p,y} + \mathtt{i}(v_{p,x} - u_{p,y})},
\end{align}
which is just a rewriting of $\mu = w_{\bar{z}}/w_{z}$ using real quantities. 

The Beltrami equation is solved numerically using the least-square method described in Section \ref{sec:num_beltrami}. The Beltrami coefficient is set to $\mu_{p}$ and the constants are fixed to $C_{u} = C_{v} = 0.5$, ensuring that $\Vert \mu_{p}\Vert_{\infty} \approx 0.121 < 1$ and that the mapping $u_{p},v_{p}$ is a quasiconformal transformation. The domain is chosen as $\Omega = [-2,2] \times [-2,2]$ and the equation discretized with a homogeneous finite element mesh of size $h$. To verify the convergence rate of the numerical method, we make $h$ vary from 0.0325 to 1.0. Usual Lagrange second order finite elements ($P^{2}$) are used.

\begin{figure}
	\begin{center}
		\includegraphics[width=0.45\textwidth]{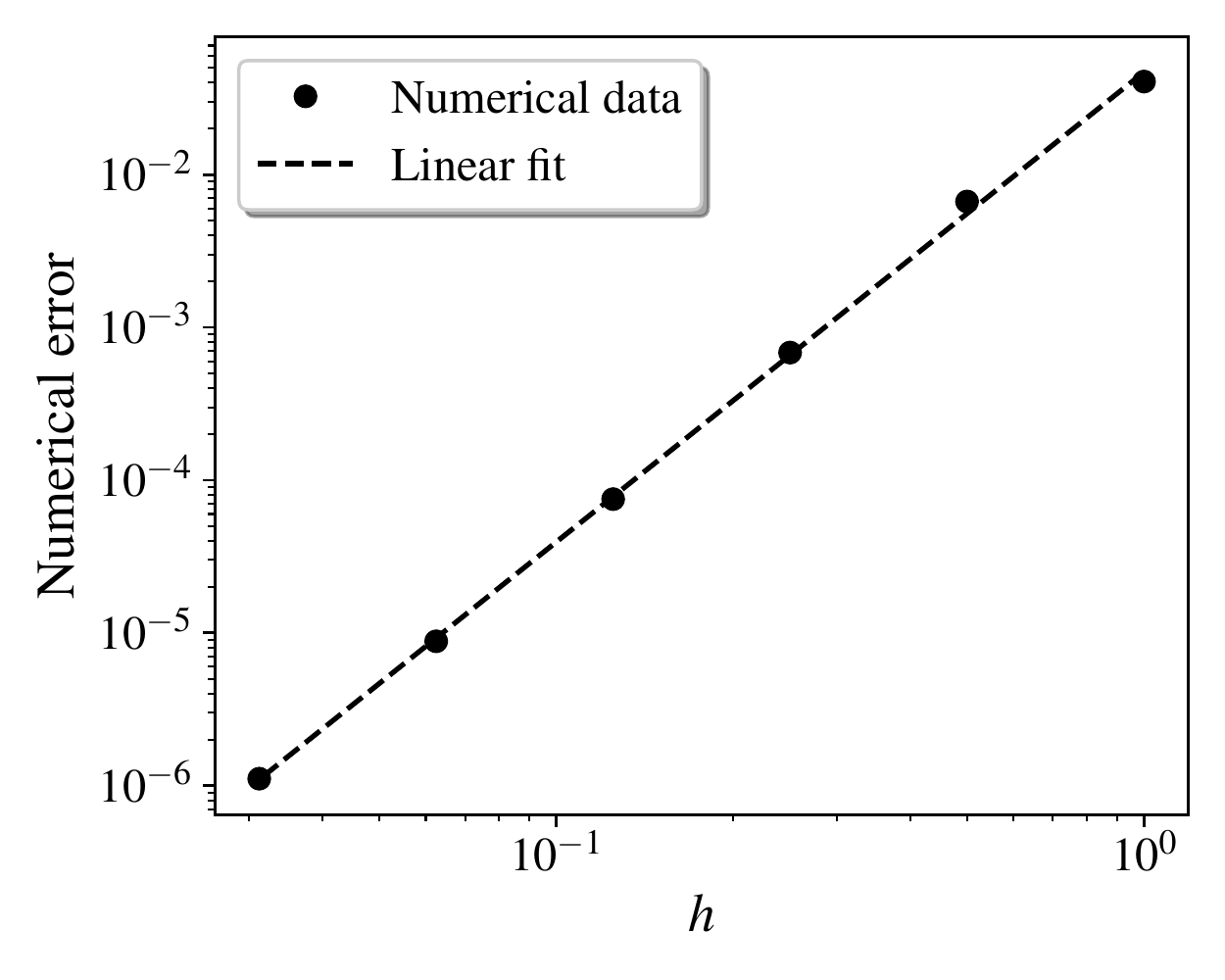}
	\end{center}
	\caption{Numerical $L^{2}$ error as a function of element size for the solution of the Beltrami equation using the least-square finite element method with P$_{2}$ elements.}
	\label{fig:conv_belt}
\end{figure}

The $L^{2}$-norm of the error is evaluated as a function of the element size and the results are displayed in Fig. \ref{fig:conv_belt}. These results demonstrate that the numerical method reproduces the exact solution with a high accuracy and that the least-square finite element method is well-suited for solving the Beltrami equation. By fitting the numerical data, we determine that the order of convergence is 3.08, as expected from the order of Lagrange polynomials used and the analysis presented in the last section.  

\subsubsection{An explicit example: the Gaussian surface}

As an illustration, a numerical experiment for the computation of $\rho({\boldsymbol u})$ is proposed.  We consider a physically relevant configuration where a graphene surface is subjected to a local Gaussian deformation with a surface $\mathcal{S}$ parameterized as follows:
\begin{align}
	X({\boldsymbol x}) &= x,  \; Y({\boldsymbol x}) = y,  \nonumber \\
	Z({\boldsymbol x})&=C\exp \left(- \frac{\vert \boldsymbol{x} \vert^{2}}{\alpha^{2}} \right), 
\end{align} 
for some given amplitude $C=20$ nm and width $\alpha= 10$ nm. The domain is $\Omega = (-50\; \mbox{nm},50 \; \mbox{nm})^2$. A deformation like this could be implemented experimentally by placing graphene on top of nanopillars or nanostructures \cite{PhysRevB.86.041405,doi:10.1063/1.5074182,Tomori_2011}.   

\begin{figure*}
	\begin{center}
		\subfloat[]{\includegraphics[width=0.9\textwidth]{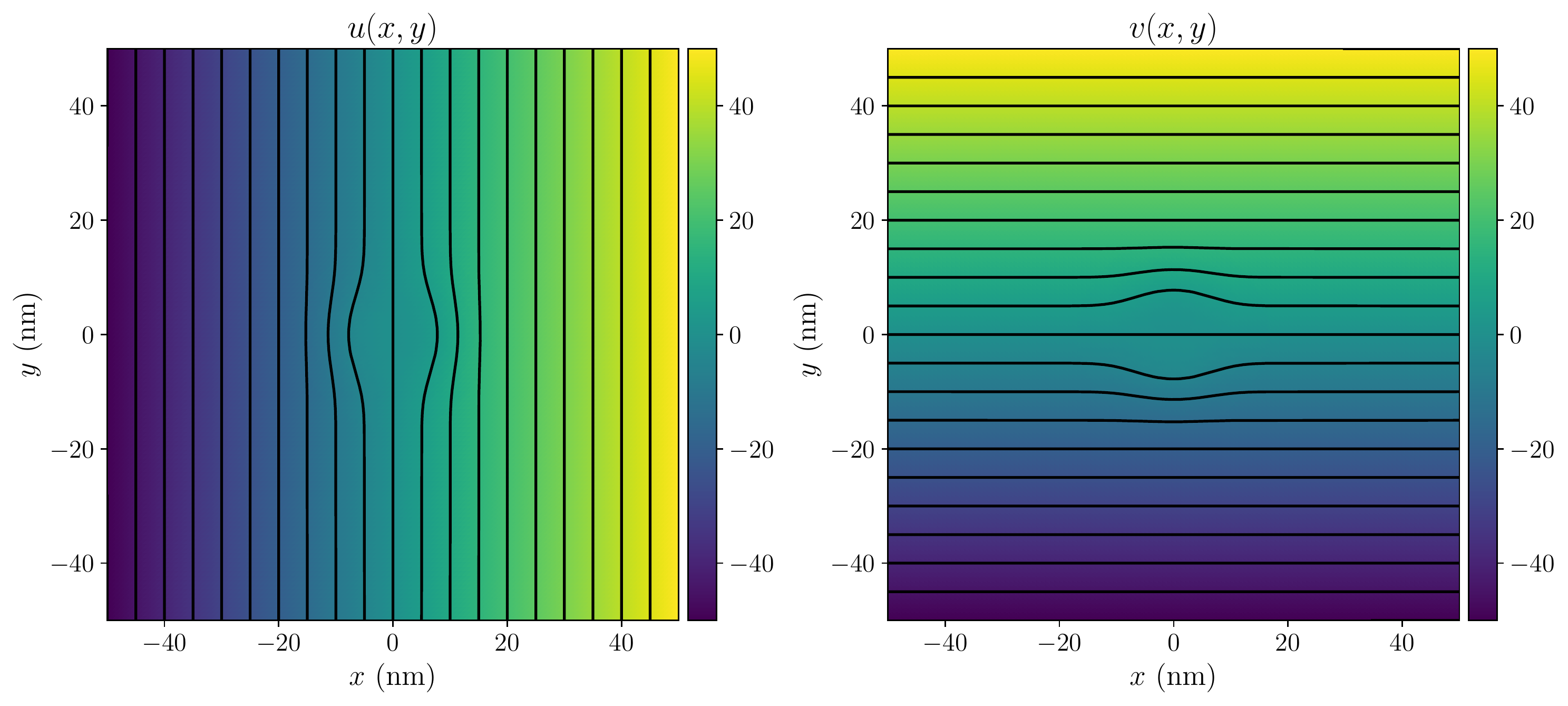}} \\
		\subfloat[]{\includegraphics[width=0.9\textwidth]{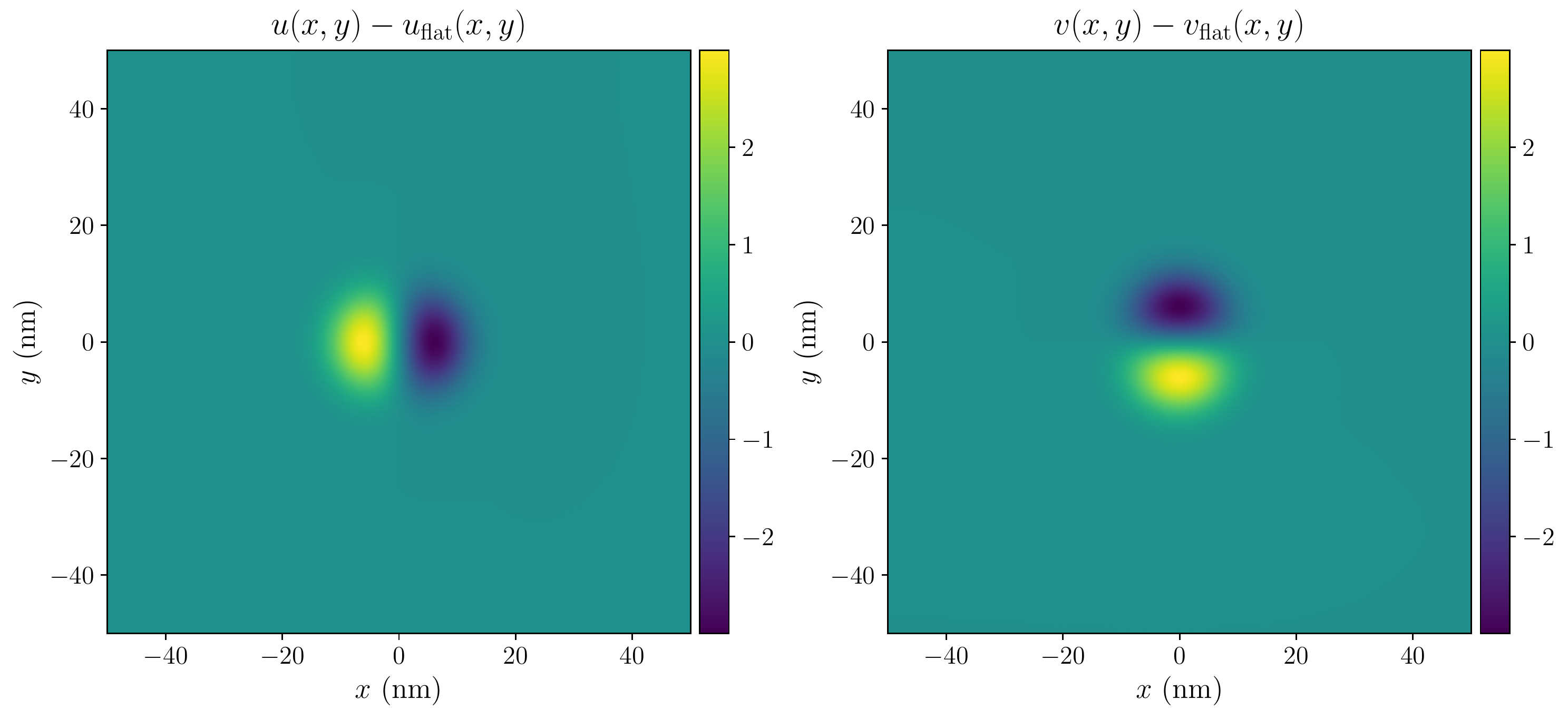}}
	\end{center}
	\caption{Graph of the (a) solution to the Beltrami equation, with $u(x,y)$ on the left and $v(x,y)$ on the right, and (b) the difference with the flat solutions $u_{\mathrm{flat}}(x,y)$ and $v_{\mathrm{flat}}(x,y)$.}
	\label{fig:sol_bel}
\end{figure*}

%
The solution to the Beltrami equation is obtained using the least square finite element numerical method described in the last section. We consider a real space triangulation of the rectangular domain with $N_x=N_y = 128$ elements of equal size in the $x$- and $y$-direction, respectively. The solution is displayed in Fig. \ref{fig:sol_bel} along with the difference with $u_{\mathrm{flat}}$ and $v_{\mathrm{flat}}$, where $u_{\mathrm{flat}},v_{\mathrm{flat}}$ are the solution to the Beltrami equation in flat space. They are obtained by setting $\mu=0$, in which case the Beltrami equation becomes the Cauchy-Riemann equation. The contours in Fig. \ref{fig:sol_bel} demonstrate clearly that the coordinates $u$ and $v$ correspond to Cartesian coordinates far from the deformation. However, they are deformed close to the region where the curvature is maximal. The difference with flat solutions allows for describing the relative effect of curvature on the Beltrami equation solution. Again, it is seen that far from the deformation, isothermal coordinates are flat while they display the effect of curvature in the vicinity of the deformation.    

\begin{figure*}
	\begin{center}
		\includegraphics[width=0.9\textwidth]{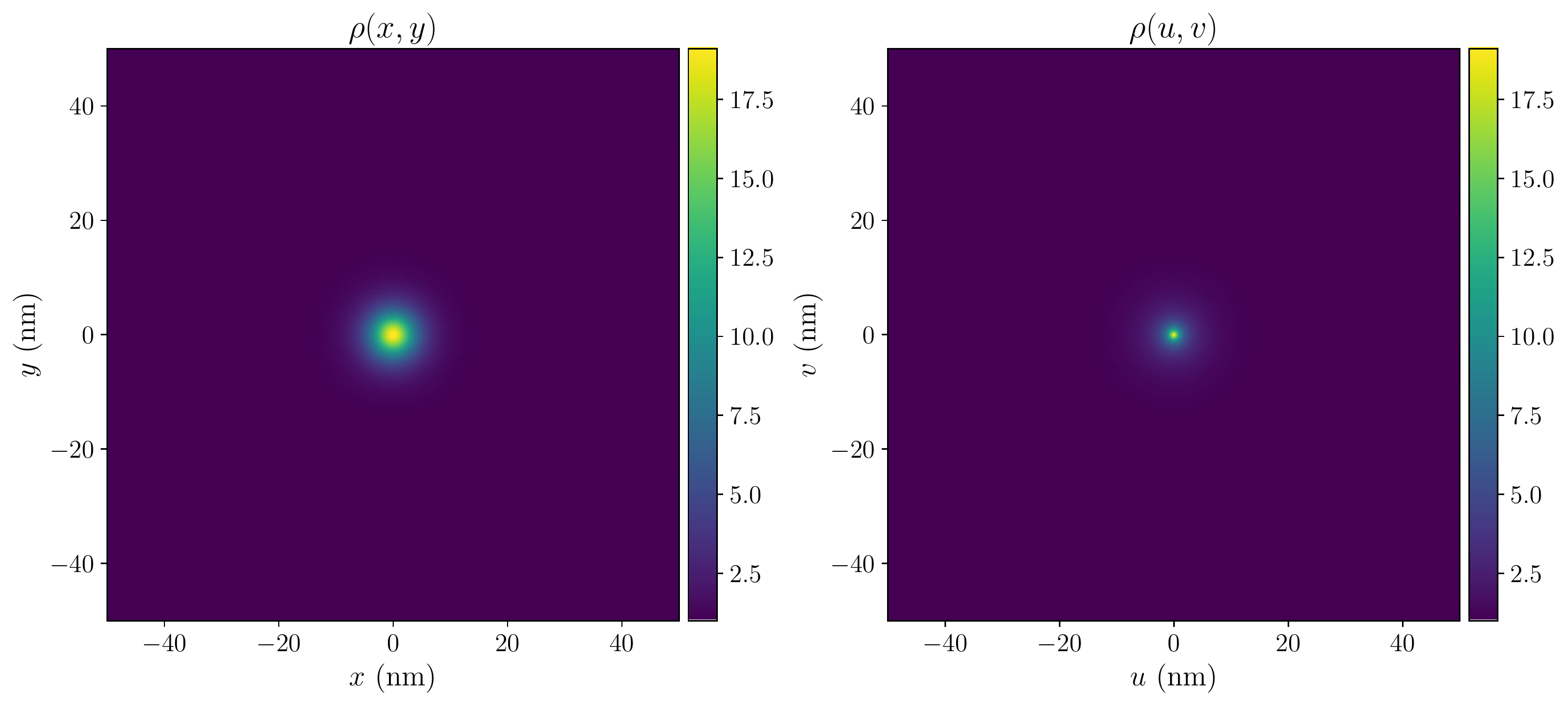}
	\end{center}
	\caption{Graph of (left) the function $\rho(\boldsymbol{x})$ for a Gaussian surface and (right) its corresponding function in isothermal coordinates $\rho(\boldsymbol{u})$.}
	\label{fig:rho}
\end{figure*}

Once $u$ and $v$ are calculated, it is possible to evaluate $\rho(\boldsymbol{x})$ via Eq. \eqref{eq:isotherm_rho}. It is displayed on the left of Fig. \ref{fig:rho}. Finally, we report in Fig. \ref{fig:rho} (Right) the function $\rho(\boldsymbol{u})$ in isothermal coordinates. The latter is obtained by interpolating the function $\rho(\boldsymbol{x})$ in Cartesian coordinates and by using the mapping $u(\boldsymbol{x})$ and $v(\boldsymbol{x})$.  As expected, we observe that $\rho(\boldsymbol{u})$ is equal to $1$ everywhere but in a small region centered at $(u,v)=(0,0)$.

\begin{figure}
	\begin{center}
		\includegraphics[width=0.45\textwidth]{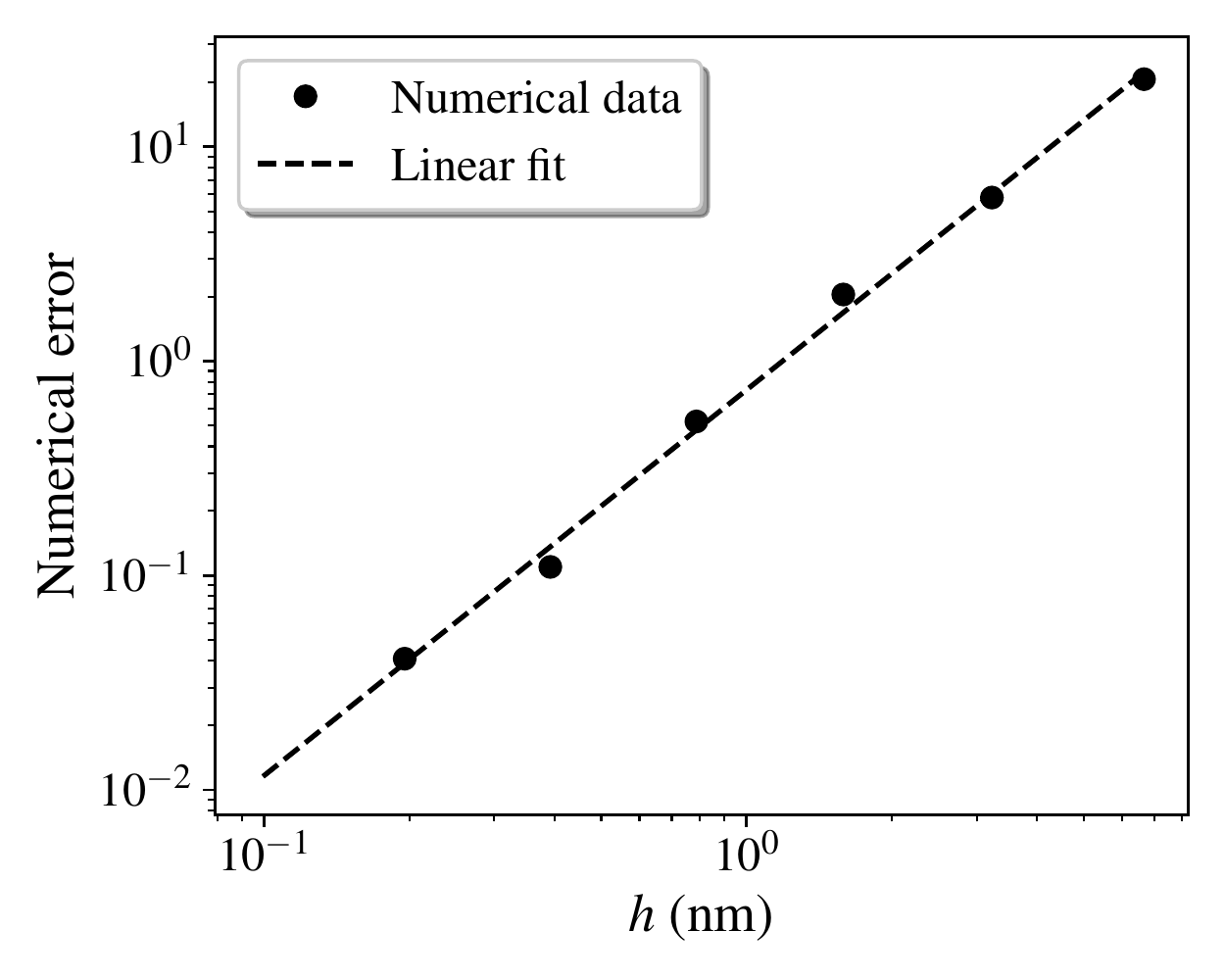}
	\end{center}
	\caption{Numerical $L^{2}$ error as a function of element size for $\rho(x,y)$ obtained from the solution of the Beltrami equation using the least-square finite element method with P$_{2}$ elements.}
	\label{fig:conv_rho}
\end{figure}

The most important function for the time-dependent solver is $\rho(\boldsymbol{x})$ given by \eqref{eq:isotherm_rho}, because it appears explicitly in the curved-space Dirac equation in isothermal coordinates.  However, this function depends on the derivative of the Beltrami solution. Here, the convergence of this function is verified as the number of elements is increased from $N_{x} = N_{y} = 16$ to $N_{x} = N_{y} = 512$. The solution of reference is approximated by setting $N_{x} = N_{y} = 1024$ and the $L^{2}$-norm of the error is calculated. The results are displayed in Fig. \ref{fig:conv_rho}. The order of convergence is numerically evaluated from the linear fit and is given by 1.799, an order lower than for the solution. This is expected because the function $\rho$ depends on the derivative of the solution. These derivatives are evaluated by taking the derivative of the polynomial basis functions, thus reducing the polynomial order. As a consequence, the order of convergence of $\rho$ is also reduced by one. This demonstrates the importance of choosing finite elements of type $P_{n}$ with $n \geq 2$ in these calculations. 

\subsection{Scattering on deformations}

This section is devoted to numerical tests and benchmarks for the pseudo-spectral method. For this purpose, we consider a simple configuration where an initial Gaussian wave packet is evolved in time and scatters on two Gaussian surface deformations. As in the last section, these local deformations could be implemented experimentally by using nanopillars or nanostructures \cite{PhysRevB.86.041405,doi:10.1063/1.5074182,Tomori_2011}. Similar configurations have been analyzed in the static regime in Refs. \cite{Milovanovi__2016,PhysRevB.88.035446,PhysRevB.90.075426} while the dynamics have been considered in Ref. \cite{PhysRevB.82.205430} for homogeneously strained graphene inducing pseudo-magnetic potentials and in Refs. \cite{Flouris2018,Debus2018} for local but symmetric deformations.

Throughout, the initial state is a Gaussian wave packet given by
\begin{align}
\psi^{0}_{+}(\boldsymbol{x}) &= \mathcal{N} e^{-\frac{\boldsymbol{x}^{2}}{\beta^{2}}}e^{i \boldsymbol{k} \cdot \boldsymbol{x}} , \\
\psi^{0}_{-}(\boldsymbol{x}) &= 0,
\end{align}
where $\beta$ is the width of the wave packet and $\boldsymbol{k}$ is its momentum. The normalization constant $\mathcal{N}$ is chosen such that the norm of the wave function is unity. The dynamics of similar wave packets  in flat graphene and the \textit{Zitterbewegung} effect have already been investigated in Ref. \cite{PhysRevB.78.235321}.   

The graphene surface, on the other hand, is parametrized by
\begin{align}
	X({\boldsymbol x}) &= x, \; Y({\boldsymbol x}) = y,  \\
	 Z({\boldsymbol x})&= \sum_{\ell = 1}^{n_{G}}C_{\ell}\exp \left(- \frac{\vert \boldsymbol{x} - \boldsymbol{a}_{\ell} \vert^{2}}{\alpha_{\ell}^{2}} \right), 
\end{align} 
corresponding to a set of $n_{G}$ Gaussian deformations, centered on $\{\boldsymbol{a}_{\ell}\}_{\ell=1,\cdots,n_{G}}$ with widths $\{\alpha_{\ell}\}_{\ell=1,\cdots,n_{G}}$ and amplitudes $\{C_{\ell}\}_{\ell=1,\cdots,n_{G}}$. 

\subsubsection{Scattering of the wave packet}
\label{sec:scat_wp}

Before investigating the properties of the numerical scheme, such as convergence, the validity of the diagonal approximation and the conservation of the norm, typical simulation results are displayed for illustration purpose and as subsequent analysis will be performed on similar configurations. The values of the simulation parameters are given in Table \ref{tab:param_sim_diff}. In addition, the domain is a square centered at the origin with sides of 200 nm, discretized with $N_{x} = N_{y} = 1024$ grid points. The Dirac equation in curved space is solved numerically using the Crank-Nicolson pseudo-spectral method in isothermal coordinates.  The solution is mapped back to Cartesian coordinates for visualization. The final time is set to $t_{\mathrm{final}} = 80$ fs and the number of time step is $N_{t} = 1000$, making for a time step $\Delta t \approx 0.08$ fs. Isothermal coordinates are obtained numerically by solving the Beltrami equation. For this purpose, the domain is discretized with a large number of elements $N_{\mathrm{el},x} = N_{\mathrm{el},y} = 1024$, ensuring that the solution is converged.

\begin{table}[t!]
	\centering
	\caption{Simulation parameters for the initial wave packet and the surface deformation for a typical scattering simulation.}
	\begin{tabular}{lc}
		\hline \hline
		Parameters & Value \\
		\hline
		Gaussian width ($\beta)$ &  10 nm \\
		Wave packet momentum $k_{x}$ & 3.12 $\times 10^{-7}$ eV \\
		Wave packet momentum $k_{y}$ & 0\\
		Number of deformation ($n_{G}$) & 2 \\
		Gaussian widths ($\alpha_{1} = \alpha_{2}$) & 10 nm\\
		Deformation position ($\boldsymbol{a}_{1}$) & (-40 nm, 0)\\
		Deformation position ($\boldsymbol{a}_{2}$) & (40 nm, 0)\\
		\hline \hline
	\end{tabular}
	\label{tab:param_sim_diff}
\end{table}

\begin{figure}
	\begin{center}
		\subfloat[$t = 80$ fs (flat)]{\includegraphics[width=0.5\textwidth]{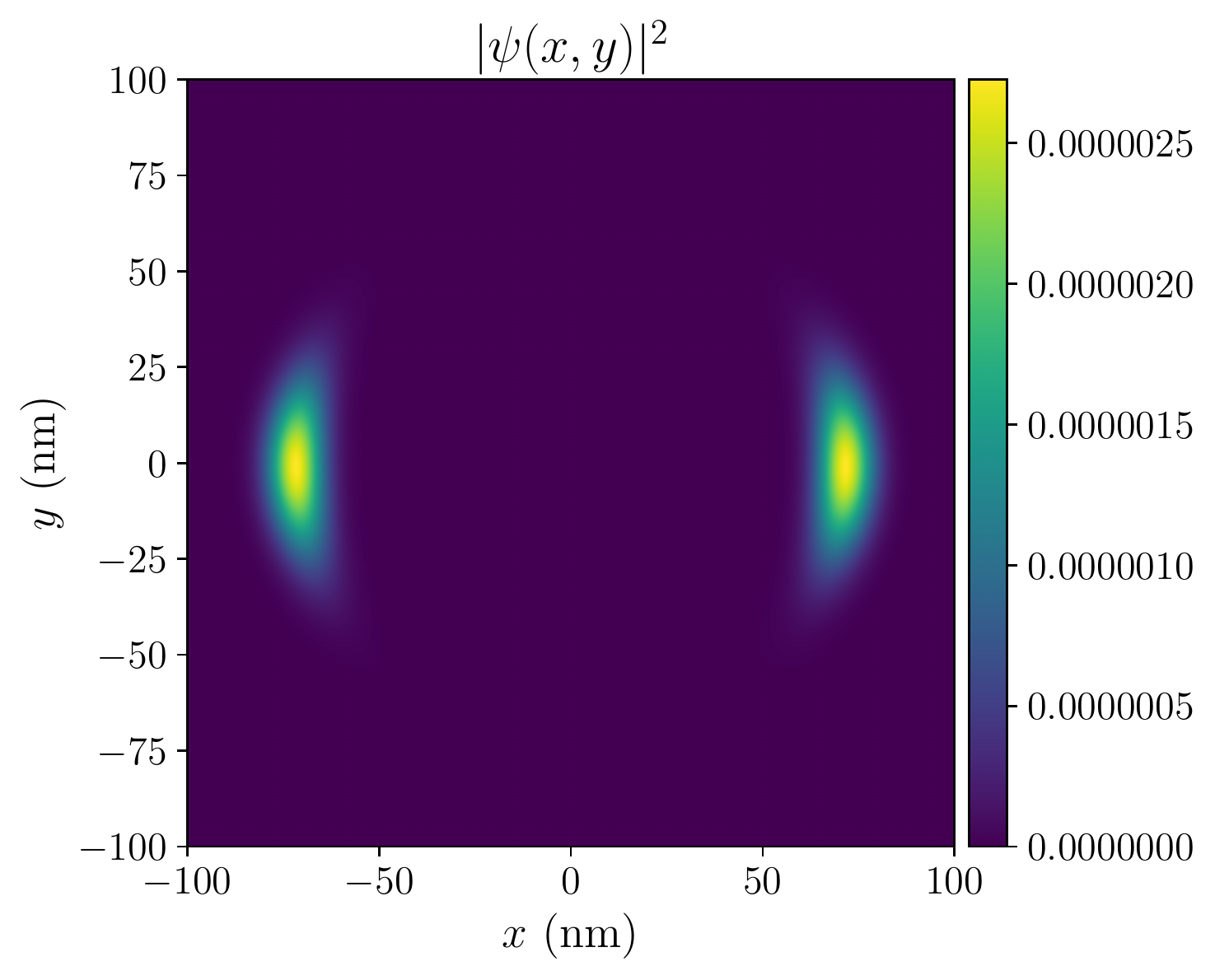}} \\
		\subfloat[$t = 80$ fs (Gaussian deformation)]{\includegraphics[width=0.5\textwidth]{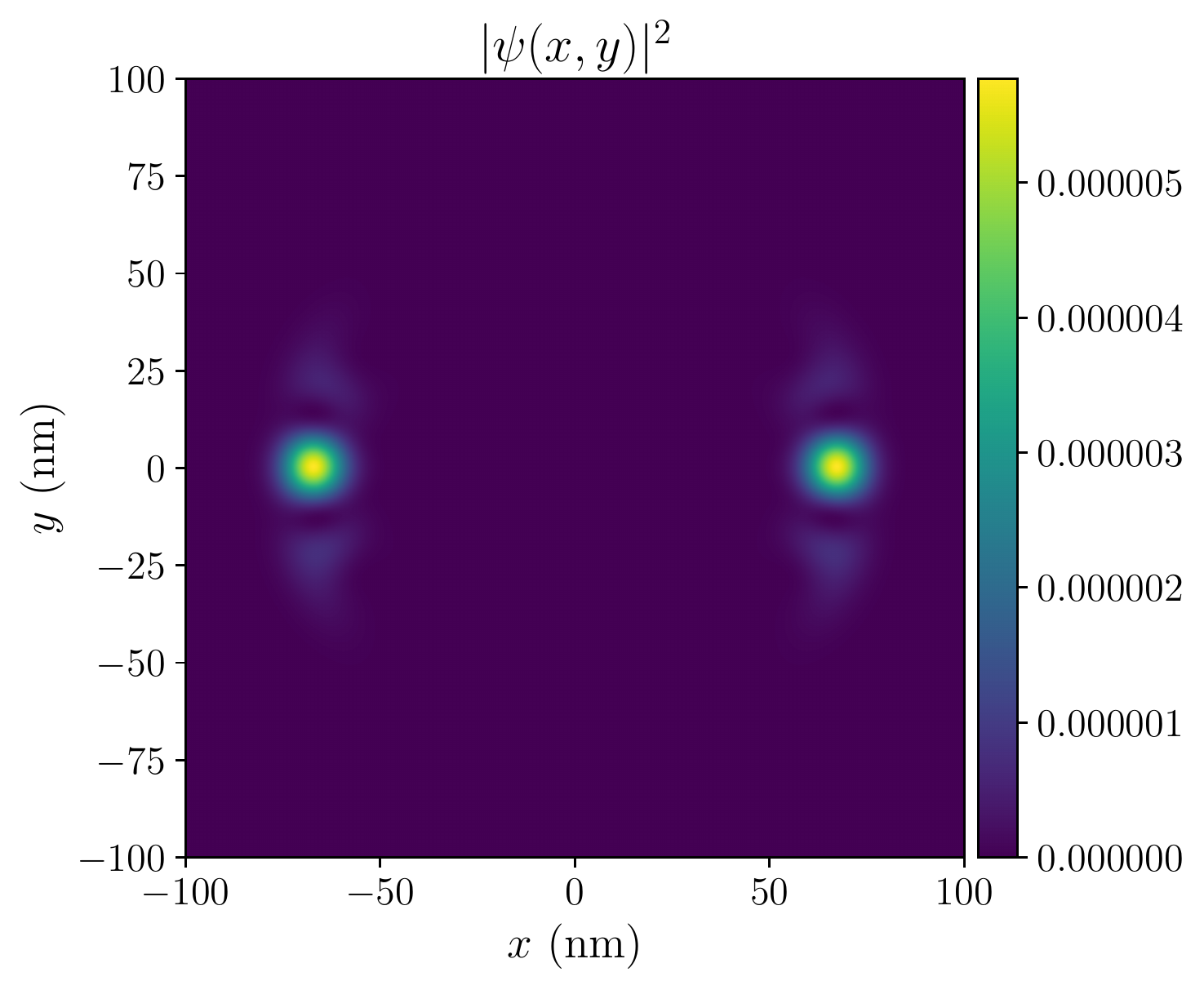}}
	\end{center}
	\caption{Plot of the wave function at final time ($t=80$ fs) when it propagates in flat space (a) and when it scatters on Gaussian deformations (b). }
	\label{fig:scat}
\end{figure}

The initial state and the final solution are displayed in Fig. \ref{fig:scat}. They are compared to the solution in flat space, where $C_{1} = C_{2} = 0$. Clearly, the presence of the deformation has an important effect on the dynamics of the wave packet, which proceeds as follow. When an initial momentum is given to the wave packet, the latter splits in two counterpropagating parts. After the free propagation in the first 50 fs, both parts reach deformed regions and scatter on Gaussian deformations. Remarkably, the scattering induces a focusing effect: the wave packet is squeezed and then starts to diverge (shown in the figure). This effect is reminiscent of gravitational lensing in general relativity, where the propagation of waves and particles is distorted by the presence of large mass objects that curve space \cite{BARTELMANN2001291}. In more details, as it propagates on deformed graphene, the center of the wave packet has to travel a longer distance than its sides because graphene is stretched more on the principal axis. The difference of propagation distances along the Gaussian deformation induces a phase difference that changes the wave front from planar to spherical and that focuses the wave packet like a lens. Analogously, the gravitational field around massive objects stretches space-time and particles that travel closer to these objects have to cover larger distances. In turns, this effect produces a wave front distortion and a focusing of particles. In both cases, the phase difference is due to the stretching of the space where particles are propagating, in contrast to an optical lens, where the phase difference comes from a modification in the optical path length. Physical implications of this electronic focusing phenomenon and its use for controlling the dynamics of electron in graphene will be investigated in more details in a subsequent article.
\begin{figure}
	\begin{center}
		\includegraphics[width=0.50\textwidth]{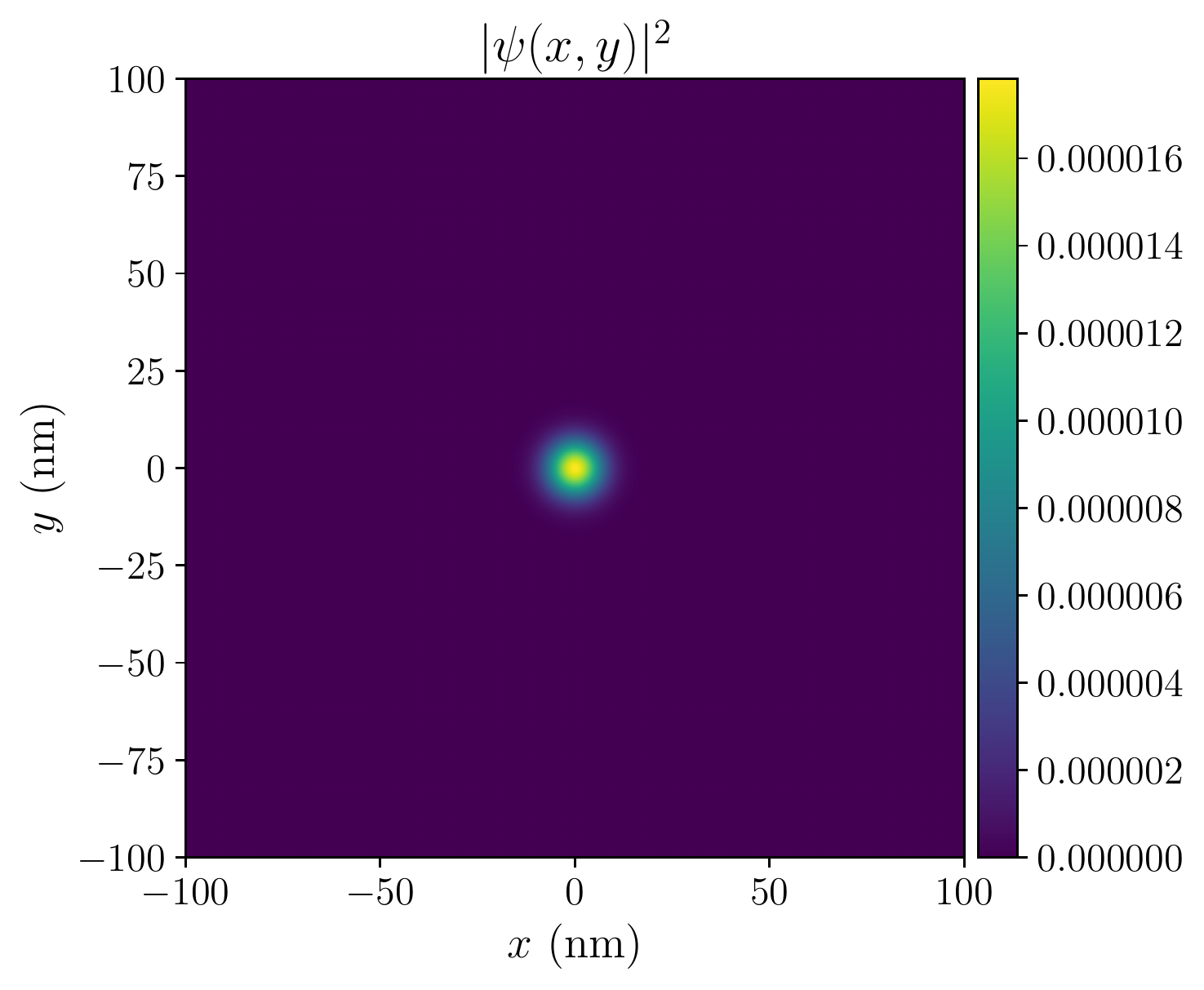} \\
	\end{center}
	\caption{Plot of the wave function at initial time $t=0$ }
	\label{fig:scat_init}
\end{figure}

\subsubsection{Diagonal approximation versus isothermal coordinates}

\begin{figure}
	\begin{center}
		\includegraphics[width=0.45\textwidth]{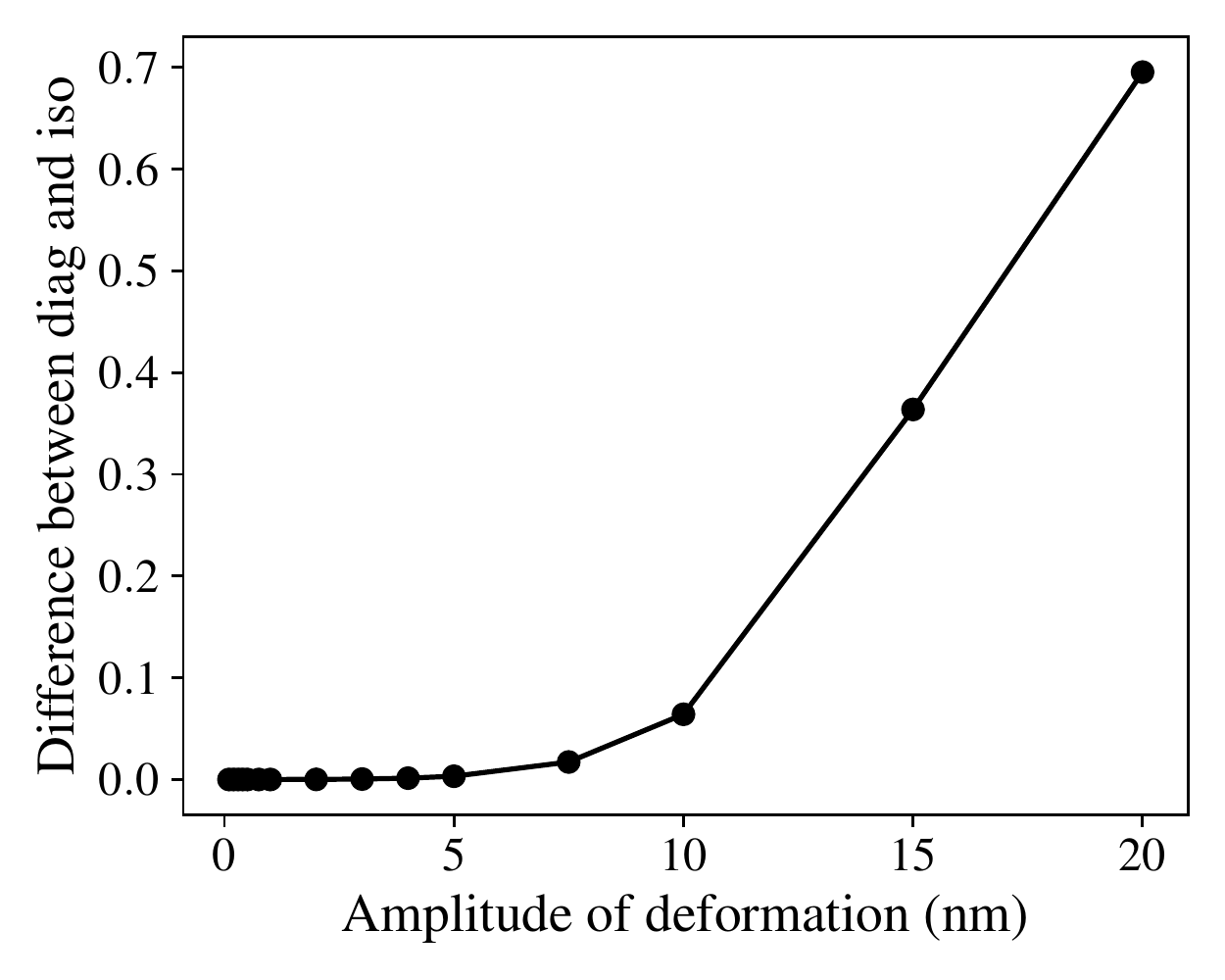}
	\end{center}
	\caption{Difference between the final solution obtained using the diagonal approximation and isothermal coordinates, as a function of the Gaussian deformation amplitude.}
	\label{fig:diff_diag_vs_iso}
\end{figure}

To empirically compare the diagonal approximation to the use of isothermal coordinates, a benchmark test is considered in which a wave packet is initialized at the center of the domain. It is given a certain momentum and thus, evolves for a small time in flat space before it is scattered on a deformation. The parameters used in simulations are the same as for Section \ref{sec:scat_wp}, except for the amplitude of the Gaussian deformation that we make vary  between 0.1 nm and 20.0 nm. Also, the Dirac equation in curved space is solved numerically using both the diagonal approximation and isothermal coordinates. The difference between the final solution in the diagonal approximation $\psi_{\eta,\mathrm{diag}}(t_{\mathrm{final}})$ and the one using isothermal coordinates $\psi_{\eta,\mathrm{iso}}(t_{\mathrm{final}})$ is quantified by evaluating the $L^{2}$-norm in Cartesian coordinates (the solution in isothermal coordinates is mapped to Cartesian coordinates): $\Delta = \frac{1}{2}\Vert \psi_{\eta,\mathrm{diag}}(t_{\mathrm{final}}) - \psi_{\eta,\mathrm{iso}}(t_{\mathrm{final}}) \Vert_{L^{2}}$. Defined in this way, $\Delta$ takes its values in the interval $\Delta \in [0,1]$, where the largest value corresponds to when the two solutions do not overlap (assuming they are normalized to 1). The result of this procedure is displayed in Fig. \ref{fig:diff_diag_vs_iso}.

As expected, the final solution in the diagonal approximation is different from the one obtained with isothermal coordinates when the deformation is larger. In our configuration, it reaches a maximum value of 0.65 when $C_{1} = C_{2} = $ 20 nm, implying that the overlap of the two functions is very small in this case. These results are expected because the non-diagonal components of the metric become important for such deformations. However, for small deformation, the difference between the two methods becomes negligible. Given that graphene can sustain a maximum strain of 25\%, it is possible that for many relevant physical configurations, the diagonal approximation may be accurate enough.

\subsubsection{Conservation of the norm in $\eta$-representation}

\begin{figure}
	\begin{center}
		\label{fig:error_norm_128}
		\subfloat[]{\includegraphics[width=0.45\textwidth]{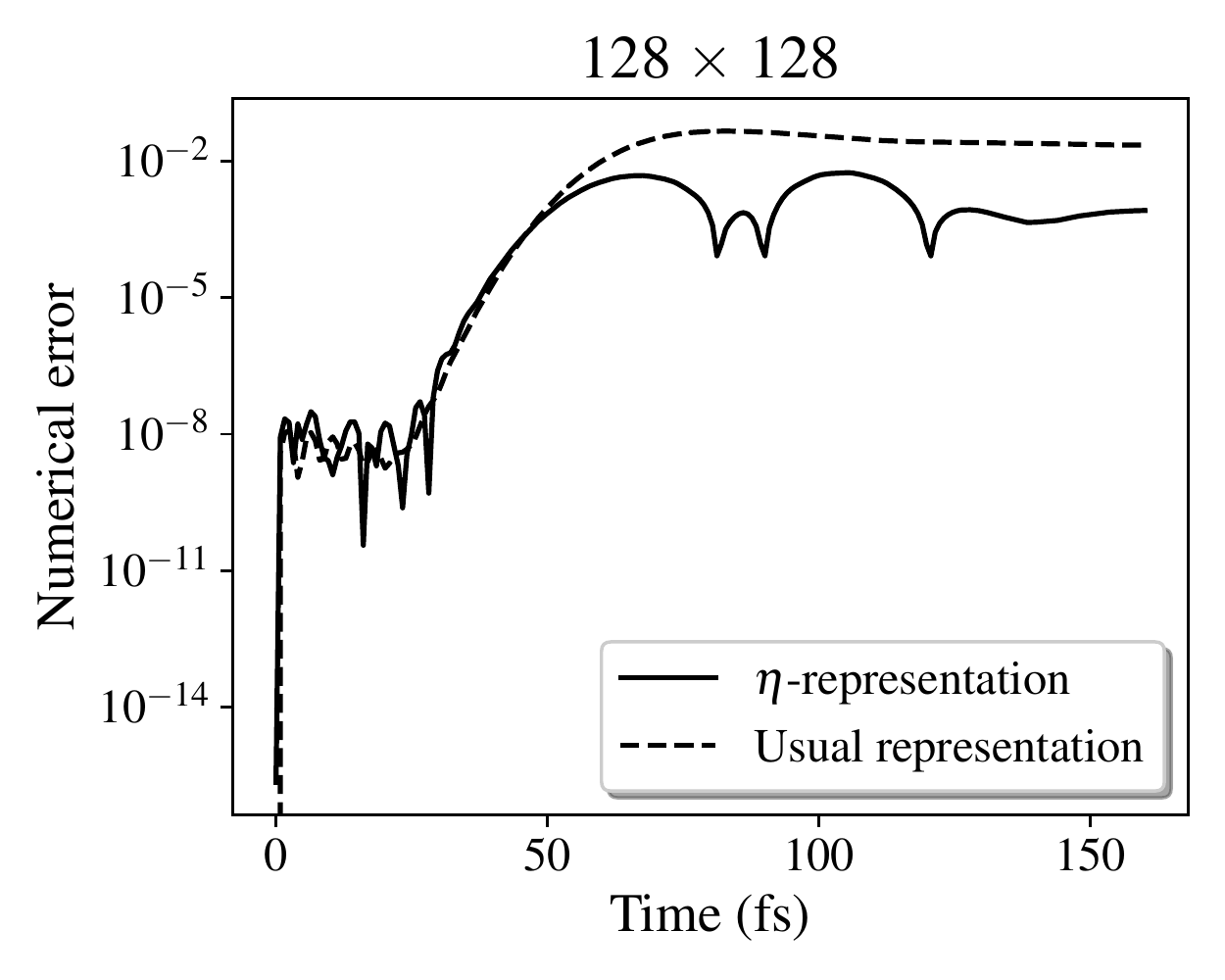}} \\
		\label{fig:error_norm_256}
		\subfloat[]{\includegraphics[width=0.45\textwidth]{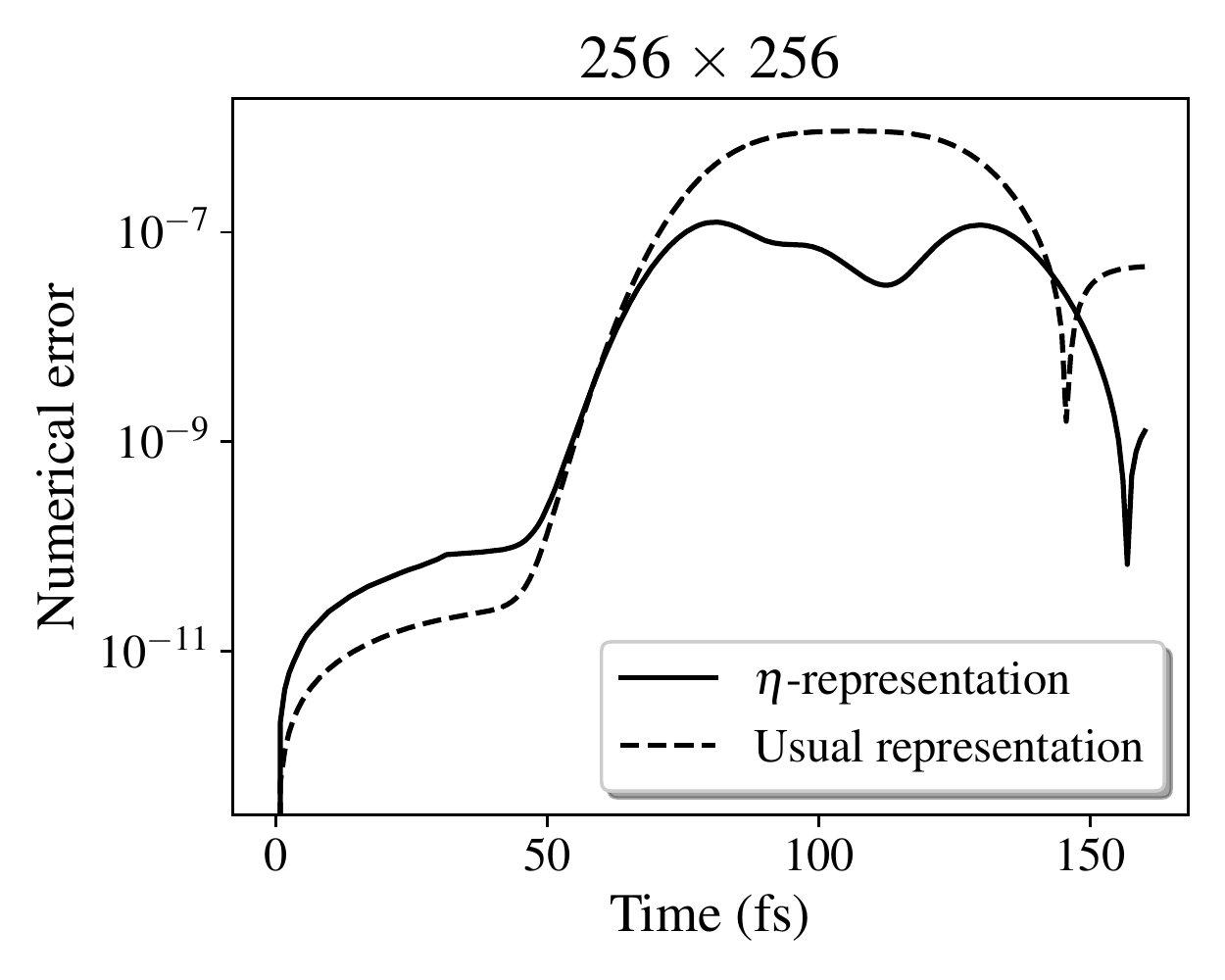}} \\
		\label{fig:error_norm_512}
		\subfloat[]{\includegraphics[width=0.45\textwidth]{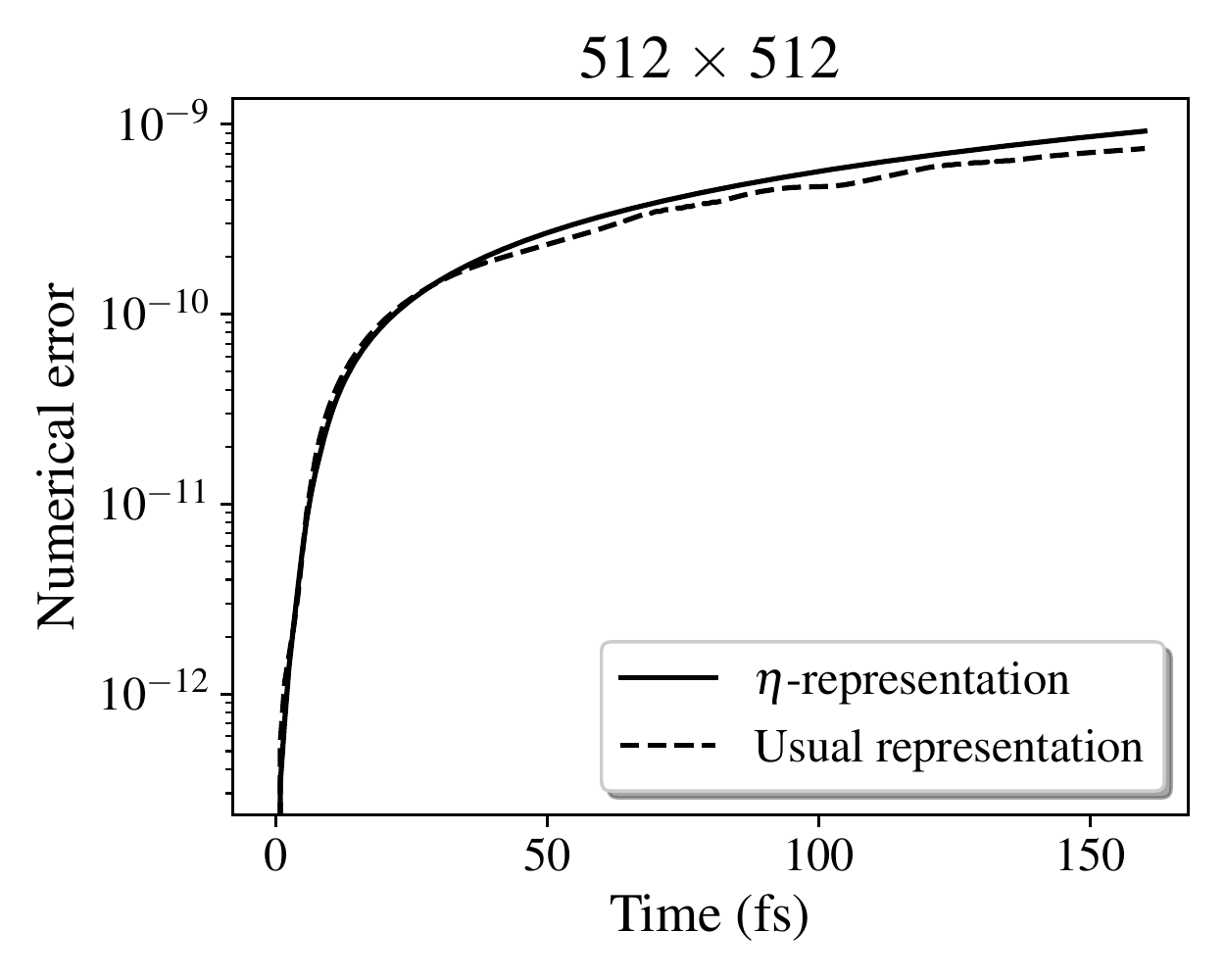}}
	\end{center}
	\caption{Numerical error of the norm as a function of time, for different grid sizes. (Top Left) $N_x\times N_y=128\times128$. (Top Right) $N_x\times N_y=256\times 256$.  (Bottom) $N_x\times N_y=512\times 512$. }
	\label{fig:error_norm}
\end{figure}

In this section, the conservation of the norm is empirically analyzed. In Section \ref{sec:dirac}, the pseudo-Hermitian Hamiltonian was transformed to the $\eta$-representation in order to ensure that the Crank-Nicolson method remains unitary at every time step and that the $L^{2}$-norm is explicitly conserved by the numerical scheme. To verify the consequence of this transformation on the conservation of the norm, we first solve the Dirac equation numerically in the usual representation $i\partial_{t} \psi = H \psi$. 
As mentioned earlier, the $g$-norm $\Vert \psi \Vert_{g}$ is conserved by the dynamics in this case. This is compared to solving the Dirac equation in the $\eta$-representation given by Eq. \eqref{eq:dirac_eta}. In this case, the $L^{2}$-norm is conserved $\Vert \psi_{\eta} \Vert_{L^{2}}$. In both cases, the pseudo-spectral Crank-Nicolson scheme is used while their corresponding norms are evaluated at every time step. The simulation parameters are given in Table \ref{tab:param_sim_norm}. 
The absolute numerical error on the norm, defined as $\epsilon = |\Vert \psi \Vert_{g} - 1|$ and $\epsilon_{\eta} = |\Vert \psi_{\eta} \Vert_{L^{2}} - 1|$ in the usual- and $\eta$-representation, respectively, is displayed in Fig. \ref{fig:error_norm} for different grid sizes: $N_{x} = N_{y} = 128, 256, 512$. All the calculations are performed in the diagonal approximation in a square domain  with 400 nm sides. The initial wave packet is evolved using 200 time steps to a final time of 160 fs.

\begin{table}[t!]
	\centering
	\caption{Simulation parameters for the initial wave packet and the surface deformation for analyzing the conservation of the norm.}
	\begin{tabular}{lc}
		\hline \hline
		Parameters & Value \\
		\hline
		Gaussian width ($\beta)$ &  10 nm \\
		Wave packet momentum $k_{x}$ & 3.12 $\times 10^{-7}$ eV \\
		Wave packet momentum $k_{y}$ & 0\\
		Number of deformation ($n_{G}$) & 2 \\
		Gaussian widths ($\alpha_{1} = \alpha_{2}$) & 20 nm\\
		Gaussian amplitudes ($C_{1} = C_{2}$) & 40 nm\\
		Deformation position ($\boldsymbol{a}_{1}$) & (-70 nm, 0)\\
		Deformation position ($\boldsymbol{a}_{2}$) & (70 nm, 0)\\
		\hline \hline
	\end{tabular}
	\label{tab:param_sim_norm}
\end{table}

In all the studied cases, the norm is accurately conserved, with numerical errors never exceeding $O(10^{-2})$. Also, the error on the norm is reduced considerably as the number of grid points is increased, as expected. The error in the $\eta$-representation is always lower than in the usual representation, especially when the surface deformations are more important. The difference naturally occurs when the wave packet reaches the region in the vicinity of the deformation, at an approximate time of $t \approx 60$ fs. At earlier times, the wave packet is essentially propagating in flat space, in which case the two representations are equivalent and give similar errors on the norm. At later times, the difference in the error between the $\eta$ and usual representation reaches at most two orders of magnitude, but is reduced further for larger grids. As a matter of fact, for the finest grid $512 \times 512$, the two representations are equivalent. At this point, the accumulated truncating errors and the numerical error of the linear solver are possibly more important than the error due to the lack of unitarity, explaining the similarity between the two representations. To conclude this study, the $\eta$-representation maintains a slight advantage over the usual representation because its norm is better conserved. However, at convergence, both representation can lead to accurate results. 

\section{Conclusion}\label{sec:conclusion}
In this paper, two numerical approaches have been proposed to solve the Dirac equation in curved space for describing the dynamics of charge carriers in corrugated graphene: the diagonal approximation and isothermal coordinates. Using these strategies, it was possible to construct a diagonal metric and as a consequence, obtain a relatively simple Dirac equation, having a very convenient form from the computational point of view. To obtain the quasi-conformal transformation allowing for a change of variable between Cartesian and isothermal coordinates, a least square numerical scheme was introduced to solve the Beltrami equation. This was tested numerically using some benchmark tests. 

The resulting Dirac equations were solved in the pseudo-Hermitian representation using the PSCN scheme. Several experiments were performed to illustrate the efficiency of the proposed methodologies for the simulation of electron dynamics on arbitrary graphene surfaces. In particular, we applied the numerical schemes to the the scattering of wave packets on local Gaussian deformations. Using these numerical tests, it was possible to conclude that the norm is better conserved in the pseudo-Hermitian representation and that the diagonal approximate ceases to be accurate for large deformations. 

The strategies presented in this paper, based on numerical quasi-conformal transformations and the diagonal approximation, could also be relevant in other contexts, for other deformed 2D physical systems or more generally, when particles are confined to move on a curved 2D plane \cite{PhysRevA.23.1982,PhysRevLett.100.230403}. Indeed, the diagonalization of the metric would also yield simpler equations in these cases described by the Schrodinger equation in curved space. Given the vast effort in 2D materials and assuming that such systems can be described by effective Schrodinger-like equations in curved space, it is plausible that our techniques could find other applications in that area. 

Physically, the results for the scattering of wave packets on local deformations have shown a focusing effect, reminiscent of gravitational lensing around massive objects in general relativity. This interesting phenomenon will be characterized in more details in a subsequent study.

\begin{acknowledgments} 
	The authors would like to acknowledge P. Levesque for many discussions on the article. This research was enabled in part by support provided by Calcul Qu\'{e}bec (www.calculquebec.ca) and Compute Canada (www.computecanada.ca).
\end{acknowledgments}

\appendix

\section{Some properties of the vielbein \label{app:prop_veil}}

The vielbein is a concatenation of three vector fields that define a local orthonormal basis. It obeys a number of important properties. First, it can be used to express the metric in the local frame fields via
\begin{align}
\label{eq:vielbein}
e_{\mu}^{A}(q) e^{B}_{\nu}(q) \eta_{AB} &= g_{\mu \nu}(q).
\end{align}
Second, it is orthonormal:
\begin{align}
e_{\mu}^{A}(q)e^{\nu}_{A}(q) &= \delta_{\mu}^{\nu} \\
e^{A}_{\mu}(q)e^{\mu}_{B}(q) &= \delta_{B}^{A} \, ,
\end{align}
where $\delta_{A}^{B}$ is the Kronecker delta.
Finally, the general and Minkowski indices can be lowered/raised using the general and Minkowski metric, respectively:
\begin{align} 
e^{\mu A}(q) &= g^{\mu \nu}(q) e^{A}_{\nu}(q), \\
e^{\mu A}(q) &= \eta^{AB} e^{\mu}_{B} (q).
\end{align}

\section{Computing the vielbein in Cartesian coordinates}
\label{app:veil_cart}

In this appendix, the vielbein is evaluated from Eq. \eqref{eq:vielbein}, re-written as a matrix equation: 
\begin{align}
g = e^{\top}\eta e,
\end{align}
where $g$, $e$ and $\eta$ are the 3-by-3 matrix representation of $g_{\mu \nu}$, $e_{\mu}^{A}$ and $\eta_{AB}$, respectively. To evaluate the vielbein, it can be useful to transform the equation in the basis of the metric eigenvectors. For this purpose, a similarity transformation is performed to diagonalize the metric:
\begin{align}
g' = P^{\top}gP = (P^{\top}e^{\top} P) \eta (P^{\top}e P) = e'^{\top} \eta e'.
\end{align}
The similarity transformation yields a new diagonal metric
\begin{align}
g'(\boldsymbol{x}) = 
\begin{bmatrix}
1 & 0 & 0 \\
0 & -\mathcal{A}^{(-)}(\boldsymbol{x}) & 0 \\
0 & 0 &  -\mathcal{A}^{(+)}(\boldsymbol{x})
\end{bmatrix},
\end{align}
where the diagonal entries are the eigenvalues of the metric $g$:
\begin{align}
\mathcal{A}^{(\pm)}(\boldsymbol{x}) = \frac{\mathrm{Tr}(g) \pm H(\boldsymbol{x})}{2},
\end{align}
where $H(\boldsymbol{x}) = \sqrt{\mathrm{Tr}(g)^{2} - 4\Delta}$. As usual, the transition matrix is constructed from the eigenvectors of $g$: 
\begin{align}
P = 
\begin{bmatrix}
1 & 0 &0 \\
0 & \cfrac{\sqrt{2}|F|}{\sqrt{H}\mathcal{W}^{(-)}} &  \cfrac{\sqrt{2}|F|}{\sqrt{H}\mathcal{W}^{(+)}} \\
0 &- \cfrac{|F|}{F}\cfrac{\mathcal{W}^{(-)}}{\sqrt{2H}} &  \cfrac{|F|}{F}\cfrac{\mathcal{W}^{(+)}}{\sqrt{2H}}
\end{bmatrix},
\end{align}
where
\begin{align}
\mathcal{W}^{(\pm)} = \sqrt{H \pm (G-E)}.
\end{align}
Once the metric has been diagonalized, it is straightforward to evaluate the vielbein. It gives
\begin{align}
e' =
\begin{bmatrix}
1 & 0 & 0\\
0 & \sqrt{\mathcal{A}^{(-)}} & 0 \\
0&0& \sqrt{\mathcal{A}^{(+)}}
\end{bmatrix}.
\end{align}
Going back in the canonical basis yields the following non-zero components:
\begin{align}
\label{eq:tetrad_mat_gen}
e^{0}_{0} &= 1 ,\\
e^{1}_{1} &=  \frac{2F^{2}}{H} \left[  \frac{\sqrt{\mathcal{A}^{(+)}}}{\mathcal{W}^{(+)2}}
	+ \frac{\sqrt{\mathcal{A}^{(-)}}}{\mathcal{W}^{(-)2}} \right], \\
e^{2}_{2} &= \frac{ \mathcal{W}^{(+)2}\sqrt{\mathcal{A}^{(+)}} + \mathcal{W}^{(-)2}\sqrt{\mathcal{A}^{(-)}}}{H} ,\\
e^{1}_{2} &= e^{2}_{1} = \frac{F}{H}\left[ \sqrt{\mathcal{A}^{(+)}} - \sqrt{\mathcal{A}^{(-)}}\right].
\end{align}
In the particular case where the surface is stretched uniformly in the $x$- and $y$-coordinates, and deformed in the $z$-coordinates, the technique used here to evaluate the vielbein yields the same results as Ref. \cite{Chaves_2014}. 

\section{About the regularity and uniqueness of solutions}
\label{app:reg_sol}

The well-posedness of the Dirac equation in curved space relies on relatively standard results in the theory of first order hyperbolic systems \cite{lefloch}.  For technical reasons, we here consider the infinite surface $\mathcal{S} =  \big\{(X(\boldsymbol{x}),Y(\boldsymbol{x}),Z(\boldsymbol{x})) \, \slash \, \boldsymbol{x} \in \R^2 \big\}$, where we assume that i) $X$, $Y$, $Z$ are smooth, and ii) $X_{i}$, $Y_{i}$, $Z_{i}$ are $L^{\infty}(\R^2)$. Under these assumptions, the Cauchy problem for the Dirac equations considered earlier \eqref{eq:dirac_H}, \eqref{eq:dirac_cart_diag} and \eqref{eq:dirac_iso} is well-posed. More specifically, we rewrite the Dirac equation in the  form
\begin{align}\label{WP1}
\begin{cases}
\partial_t \psi(t,{\boldsymbol q})  =   A({\boldsymbol q})\partial_1\psi(t,{\boldsymbol q})+B({\boldsymbol q})\partial_2\psi(t,{\boldsymbol q})  + C({\boldsymbol q}) \psi(t,{\boldsymbol q}) , \\
\psi(0,\cdot)  =  \psi_0:=(\psi_{+}^0,\psi_{-}^0)^T \, ,
\end{cases}
\end{align}
where $A,B,C$ are matrices related to the vielbein and the affine spin connection. They are given by
\begin{align}
A({\boldsymbol q}) &= -  \biggl\{ e_1^{1}({\boldsymbol q})\alpha^1 + e_2^{1}({\boldsymbol q})\alpha^2 \biggr\} ,\\
B({\boldsymbol q}) &= -  \biggl\{ e_1^{2}({\boldsymbol q})\alpha^1 + e_2^{2}({\boldsymbol q})\alpha^2  \biggr\}, \\
C({\boldsymbol q}) &= -  \biggl\{ \big(e_1^{1}({\boldsymbol q})\alpha^1 + e_2^{1}({\boldsymbol q})\alpha^2\big)\Omega_1({\boldsymbol q})  \nonumber \\
& +  \big(e_1^{2}({\boldsymbol q})\alpha^1 + e_2^{2}({\boldsymbol q})\alpha^2 \big)\Omega_2({\boldsymbol q}) \biggr\}.
\end{align}
A theorem of existence and uniqueness can then be stated for equations of this form \cite{existence1, existence2}. Denoting by $C_b^s$ the set of $s$ times continuously differentiable (matrix) functions with bounded derivatives, we have
\begin{theo}
	Assume that $A$, $B$, and $C$ belong to $C^{s+1}_b(\R^2)$, with $s>2$ and that $\psi_0\in H^{s}(\R^2)$, then there exists a unique solution $\psi$ to \eqref{WP1} which belongs to $\psi \in C^0\big([0,\infty);H^s(\R^2)\big)\cap C^1\big([0,\infty);H^{s-1}(\R^2)\big)$.
\end{theo}
This theorem guides the development of numerical methods and ensure that the latter will converge towards the solution.

\bibliography{refs}

\end{document}